\documentclass{aastex63}
\shorttitle{icy shell structure with admittance}
\shortauthors{Akiba et al.}
\usepackage{amsmath}
\usepackage{lineno}

\newcommand{\thickhline}{%
    \noalign {\ifnum 0=`}\fi \hrule height 1pt
    \futurelet \reserved@a 
}

\begin{document}

\title{Probing the icy shell structure of ocean worlds with gravity-topography admittance}

\correspondingauthor{Ryunosuke Akiba}
\email{ryusterakiba@berkeley.edu}
\author[0000-0002-2681-3195]{Ryunosuke Akiba}
\affiliation{Department of Earth and Planetary Science, University of California, Berkeley, CA, 94720, USA}
\author[0000-0002-7020-7061]{Anton I. Ermakov}
\affiliation{Department of Earth and Planetary Science, University of California, Berkeley, CA, 94720, USA}
\author[0000-0002-7092-5629]{Burkhard Militzer}
\affiliation{Department of Earth and Planetary Science, University of California, Berkeley, CA, 94720, USA}
\affiliation{Department of Astronomy, University of California, Berkeley, CA, 94720, USA}
\begin{abstract}

    The structure of the icy shells of ocean worlds is important for understanding the stability of their underlying oceans as it controls the rate at which heat can be transported outward and radiated to space. Future spacecraft exploration of the ocean worlds (e.g., by NASA's Europa Clipper mission) will allow for higher-resolution measurements of gravity and shape than currently available. 
    
    In this paper, we study the sensitivity of gravity-topography admittance to the structure of icy shells in preparation for future data analysis. An analytical viscous relaxation model is used to predict admittance spectra given different shell structures determined by the temperature-dependent viscosity of a tidally heated, conductive shell. We apply these methods to the ocean worlds of Europa and Enceladus. We find that admittance is sensitive to the mechanisms of topography support at different wavelengths and estimate the required gravity performance to resolve transitions between these mechanisms. We find that the Airy isostatic model is unable to accurately describe admittance universally across all wavelengths when the shell thickness is a significant fraction of body's radius. Our models suggest that measurements of admittance at low spherical harmonic degrees are more sensitive to thick shells with high tidal dissipation, and may complement ice-penetrating radar measurements in constraining shell thickness. Finally, we find that admittance may be used to constrain the tidal dissipation within the icy shell, which would be complementary to a more demanding measurement of the tidal phase lag.
    
\end{abstract}

\section{Introduction}
The existence of subsurface oceans within icy satellites is of great interest due to their potential habitability. In this paper, we focus on two such ocean worlds: Jupiter’s moon Europa and Saturn’s moon Enceladus. The presence of a global subsurface ocean on Europa has been inferred from observations by \textit{Galileo} of an induced magnetic field. A global salty ocean is believed to be the conductive fluid causing the induced magnetic field \citep{khurana1998induced}. The presence of liquid water in Enceladus was inferred from observations of water vapor plumes \citep{porco2006cassini}. \cite{thomas2016enceladus} inferred the global nature of the ocean from observations of the large amplitude of physical libration. 

Considering life as we know it, the presence of large amounts of liquid water, essential elements from chondritic material at the base of the ocean, energy from tides and radiogenic  sources as well as chemical gradients make ocean worlds promising astrobiology targets \citep{hand2009astrobiology}. A major consideration for assessing habitability is the persistence and stability over geologic timescales of a liquid water ocean to allow for developing and sustaining life. The structure of the overlying icy shell, its thickness, heat transport mechanism, and the spatial distribution of tidal heating are particularly important for understanding the overall heat budget and the long-term survivability of the ocean. 

For Europa, the structure of the icy shell has been studied using classical Airy isostasy model and observations of topography \citep[e.g.,][]{kadel2000geological}, flexural analysis \citep[e.g.,][]{billings2005great,nimmo2011geophysical}, and tidal-convective equilibrium model \citep[e.g.,][]{moore2006thermal}, but large uncertainty remains for the mean shell thickness. For Enceladus, shell thickness was constrained by gravity, shape and libration data \citep[e.g.,][]{hemingway2019enceladus}. However, for both Europa and Enceladus the properties of the icy shell that govern heat transport, such as viscosity and temperature profile, remain poorly constrained.


In this paper, we explore the use of gravity-topography admittance to study the icy shells of ocean worlds Europa and Enceladus. Gravity-topography admittance $Z_n$ is defined as a wavelength-dependent ratio of gravity to topography spectral amplitudes:

\begin{equation}
    Z_n = \frac{\mathrm{gravity \hspace{0.1 cm} amplitude}}{\mathrm{shape \hspace{0.1 cm} amplitude}} \hspace{0.1 cm} \mathrm{at \hspace{0.1 cm} spherical \hspace{0.1 cm} harmonic \hspace{0.1 cm} degree}  \hspace{0.1 cm} n \hspace{0.1 cm} [\mathrm{mGal/km}],
\end{equation}

\noindent where gravity and shape data are obtained from spacecraft observations. Gravity-topography admittance can be computed from the spherical harmonic expansion coefficients of the gravity and shape up to the degree of the lower-resolution data set. Typically, the accuracy and resolution of admittance are limited by the quality of the gravity data set. 

Current observations of gravity and shape of Europa and Enceladus from the \textit{Galileo} and \textit{Cassini} spacecraft, respectively, are limited to long-wavelength measurements. In terms of spherical harmonic expansions, gravity fields have been measured up to spherical harmonic degree~2 for Europa (spatial scale of 4900~km)
\citep{anderson1998europa,casajus2020updated} and up to degree~3 for Enceladus \citep{iess2014gravity} (spatial scale of 528~km). \cite{nimmo2007global} made ellipsoidal fits of Europa's shape, and Enceladus' shape has been mapped up to spherical harmonic degree~16 \citep{tajeddine2017true} (spatial scale of 100~km). Future missions such as the upcoming NASA's Europa Clipper will deliver higher-resolution measurements of the gravity and shape allowing determination of gravity-topography admittance over a larger range of spatial scales. Our goal is to prepare for these future higher-resolution data by exploring the sensitivity of gravity-topography admittance to various aspects of the icy shell structure.

Gravity-topography admittance bears clues to the icy shell structure, specifically to topography support mechanisms. The dominant topography support mechanism could vary depending on the spatial scale. The Airy isostasy model, used extensively for the Earth \citep[e.g.,][]{watts2001isostasy}, describes surface topography supported by buoyancy force arising due to a topography of a density-discontinuity interface (i.e., a crustal root) located at a certain depth called ``depth of compensation''. Topographic loads can also be supported by elastic and viscous stresses within the body. Since the interface between the icy shell and ocean is a phase boundary, melting and freezing can produce non-hydrostatic basal topography inducing a flow throughout the shell \citep{vcadek2019airy}. These different topography support mechanisms influence the amplitudes of topography at different wavelengths at the surface and the base of the shell, which affects the moon's gravity field and, therefore, is reflected in the admittance spectrum. 

In summary, the goals of the paper are:

\begin{enumerate}
    \item To provide an algorithm for computing gravity-topography admittance suitable for icy shells with large gradients of viscosity.
    
    \item To explore the sensitivity of gravity-topography admittance to viscosity profile, shell tidal heating, and shell thickness. 
\end{enumerate}

\section{Methods}

\begin{figure}
    \centering
    \includegraphics[width=\columnwidth]{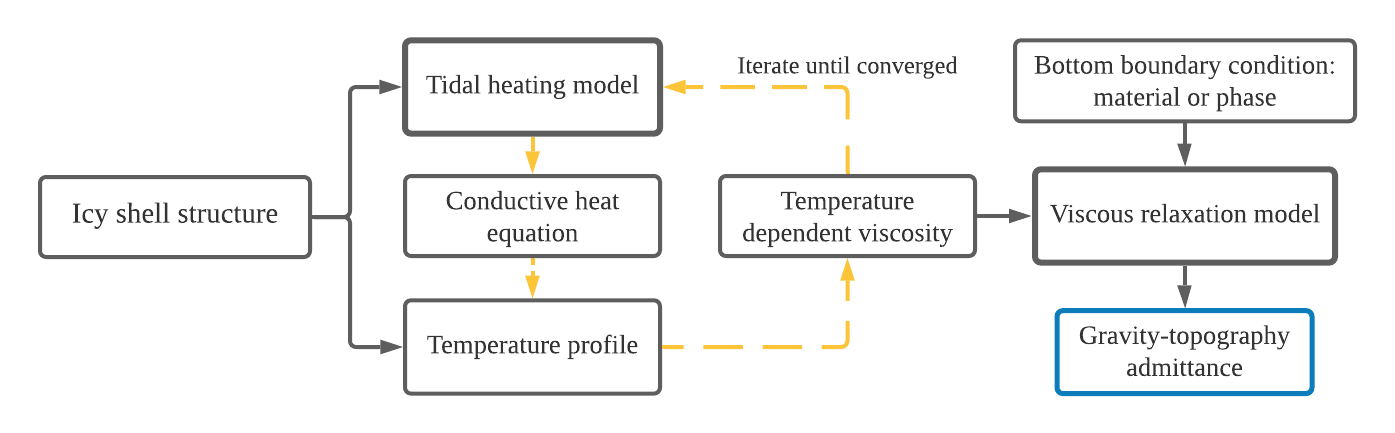}
    \caption{Flowchart for our exploration of icy shell structure.}
    \label{fig. flowchart}
\end{figure}

In our exploration of the sensitivity of admittance to icy shell structure, we follow the process laid out in Fig.~\ref{fig. flowchart}. We model Europa and Enceladus as three spherically symmetric layers, placing an icy shell on top of a liquid water ocean that overlies a solid mantle. For each choice of shell thickness, the density of the mantle and thickness of the ocean are computed to satisfy the total mass, radius, and moment of inertia factor of the moon as given in Table~\ref{tab. params}. In addition, viscoelastic parameters shown in Table~\ref{tab. params} are assigned to each layer and are used in the tidal heating model. Starting with this icy shell structure, a temperature profile is found by solving the conductive heat equation with tidal heating as a source term. Since tidal heating depends on a temperature-dependent ice rheology, we compute the temperature profile iteratively by alternating calculations of tidal heating and heat conduction to arrive at a steady-state solution. We use this converged temperature profile to determine the shell's viscosity profile. The viscosity profile is then used in a viscous relaxation model to find the shape amplitudes at the surface and the base of the icy shell, treating the ice-ocean boundary as either a material or a phase boundary. Finally, the relation between the shell's surface and base amplitude at each spherical harmonic degree is used to obtain the expected gravity-topography admittance spectrum. We will now proceed with a detailed description of these steps.

\begin{table}[htbp]
  \centering
    \begin{tabular}{lcc}
    \hline
    Parameter & \multicolumn{1}{l}{Europa} & \multicolumn{1}{l}{Enceladus}\\
    \hline
    \hline
    Radius (km) & \multicolumn{1}{r}{$1560.8 \pm 0.3$ \citep{nimmo2007global}} & \multicolumn{1}{r}{$252.24\pm 0.2$ \citep{thomas2016enceladus}}\\
     \hline
    $G M$ ($\mathrm{km}^3 \mathrm{s}^{-2}$)  & \multicolumn{1}{r}{$3202.72\pm 0.05$ \citep{anderson1998europa}} & \multicolumn{1}{r}{$7.2096\pm 0.0067$  \citep{jacobson2006gravity}} \\
     \hline
    Moment of inertia factor, $C/MR^2$ & \multicolumn{1}{r}{$0.346\pm 0.005$ \citep{anderson1998europa}} & \multicolumn{1}{r}{$0.3305 \pm 0.0025$ \citep{mckinnon2015effect}} \\
     \hline
    Eccentricity  & \multicolumn{1}{r}{0.009} & \multicolumn{1}{r}{0.0047} \\
     \hline
    Orbital period (hours) & \multicolumn{1}{r}{85.2} & \multicolumn{1}{r}{32.9} \\
    \hline
    \hline
    Ice shell density (kg $\mathrm{m}^{-3}$) & \multicolumn{2}{c}{920}\\
     \hline
    Ocean density (kg $\mathrm{m}^{-3}$) & \multicolumn{2}{c}{1050} \\
     \hline
    Poisson's ratio of shell & \multicolumn{2}{c}{0.33} \\
     \hline
    Poisson's ratio of mantle & \multicolumn{2}{c}{0.33} \\
     \hline
    Shear modulus of shell (GPa) & \multicolumn{2}{c}{3.3} \\
     \hline
    Shear modulus of mantle (GPa) & \multicolumn{2}{c}{40} \\
     \hline
    Bulk modulus of ocean (GPa) & \multicolumn{2}{c}{2.15}\\
    \hline
    Viscosity of mantle (Pa s) & \multicolumn{2}{c}{$\infty$} \\
    \hline
    \end{tabular}%
   \caption{Model Parameters. We use the central values in our models.}
   \label{tab. params}%
\end{table}%

\subsection{Assumptions of ice rheology}
\label{sec. rheology}

The rheologic model of ice governs its response to applied stresses. We assume that the icy shell flows viscously on geologic timescales. The properties of ices under conditions relevant to icy moons are poorly understood, largely due to the difficulty of reproducing low stress, frequency, and temperature conditions with laboratory experiments ($\leq 0.1$ MPa, \cite{tobie2003tidally}). The flow mechanisms of diffusion creep and grain boundary sliding (GBS) are most relevant for the low stress and strain rates, low temperatures, and small grain sizes expected in icy moons \citep{goldsby2001superplastic}. The diffusion creep flow mechanism, extrapolated to these low stresses from experimental data and described by \cite{goldsby2001superplastic}, results in a Newtonian flow, in which strain rate is proportional to stress. GBS has lower activation energy compared to diffusion creep and, unlike diffusion creep, is non-Newtonian, which introduces a stress dependence of viscosity. Diffusion creep likely dominates at the conditions and grain sizes in the icy shell \citep{moore2006thermal}. Thus, we assume viscosity is independent of stress and the flow is Newtonian described by the diffusion creep deformation mechanism \citep[e.g.,][]{showman2004numerical, tobie2003tidally,mitri2005convective}. Under these assumptions, the temperature-dependent viscosity of pure water ice is described by:

\begin{equation}
    \eta(T) = \eta_{\rm{melt}} \exp\left( \frac{E_a}{R T_m} \left(\frac{T_m}{T} - 1\right)\right),
    \label{eq. viscosity}
\end{equation}

\noindent where $E_a$ is the activation energy of diffusion creep taken to be 59.4~kJ $\mathrm{mol}^{-1}$ \citep{goldsby2001superplastic}, $T_m$ is the melting temperature of ice taken to be 273~K, and $R$~=~8.314~J mol$^{-1}$~K$^{-1}$ is the universal gas constant. The grain size dependency is included within the viscosity at the melting point $\eta_{\rm{melt}}$. The grain size is estimated to be in the range from 0.1 to 1~mm \citep{mckinnon1999convective,kirk1987thermal}, corresponding to $\eta_{\rm{melt}}$ between $10^{13}$~Pa~s and $10^{15}$~Pa~s.

There are several simplifications in our relaxation modeling. We do not model solid-state convection within the shell. Convection, if it occurs, is most likely confined to the bottom part of the shell and is less likely to occur in thinner shells \citep[e.g.,][]{mckinnon1999convective}. The inferred large amplitude of basal shell topography at Enceladus argues against convection as convection would likely rapidly relax this topography \citep{hemingway2019enceladus}. In addition, the rheology of ice at the base of the shell near the melting point can be influenced by premelting and partial melting that would reduce viscosity enhancing the flow \citep{tobie2003tidally}. Additionally, impurities within the ice, such as salts and silicates, can affect grain sizes and increase viscosity \citep{barr2009heat}. 

At the surface, applying Eq. \ref{eq. viscosity} at the equilibrium surface temperatures of Europa (92~K) or Enceladus (59~K) implies that ice near the surface has a viscosity on the order of $10^{30}$ and $10^{50}$~Pa~s, respectively. Such high viscosities will prevent the relaxation of surface topography on geologic timescales. \cite{passey1983viscosity} and \cite{bland2012enceladus} find that craters on Enceladus are highly relaxed despite these low surface temperatures. \cite{passey1983viscosity} determine a range of surface viscosities on Enceladus between $10^{24}$ and $10^{25}$~Pa~s to explain the relaxation of the observed craters, and suggest that an insulating layer at the surface could increase the effective surface temperature and thus decrease viscosity. On Europa, \cite{showman2004numerical} argue that fractures in ice and brittle deformation could affect ice rheology at the surface and parametrize this more complex rheology by also imposing an upper viscosity bound. We follow this simplification of a bounded viscosity used by \cite{showman2004numerical} and \cite{vcadek2017viscoelastic} for our viscous relaxation model and consider values $\eta_{\rm{bound}} = 10^{22} - 10^{25}$~Pa~s. We compute tidal heating (Section \ref{sec. tidal}) and the temperature structure of the shell (Section \ref{sec. temperature}) using unbounded viscosity, and the resulting viscosity profile will subsequently be set to $\eta_{\rm{bound}}$ where $\eta(r)>\eta_{\rm{bound}}$ to compute the relaxation of the shell (Section \ref{sec. relaxation}).

\subsection{The effect of tidal heating}
\label{sec. tidal}

As an icy moon orbits its parent planet, tidal forces cause a periodic deformation of the moon resulting in heat production. In this paper, we introduce tidal heating within the icy shell through its influence on the temperature profile and, thus, viscosity profile through Eq. \ref{eq. viscosity}. We use the Maxwell rheology to model viscoelastic deformation of a spherically symmetric, layered body under the influence of tidal potential. For a chosen shell thickness and viscosity profile, our layered model moon is described by the viscoelastic and orbital parameters listed in Table \ref{tab. params}. While the mantle and ocean are treated as single layers with constant viscoelastic parameters, the icy shell is broken up into 80 layers prescribed by the viscosity profile. The degree~2 tidal potential expanded to first order in eccentricity for a tidally-locked satellite is used to set the surface boundary condition. This potential drives the deformation of the viscoelastic shell causing tidal heating. We follow \cite{tobie2005tidal} to compute a radial profile of surface averaged volumetric tidal dissipation rate, $h_{\rm{tide}}(r)$. We use $h_{\rm{tide}}(r)$ as a source term in the heat conduction equation in the following Section \ref{sec. temperature}. A full description of the tidal dissipation model and the computation of $h_{\rm{tide}}(r)$ is given in Appendix \ref{apdx tidal}.

Tidal dissipation is maximized in the portion of the shell where Maxwell time, defined as the ratio of viscosity to elastic shear modulus, is close to the forcing period \citep[e.g., ][]{tobie2003tidally}. The viscosity for which this occurs can be expressed as $\eta_{\rm{Maxwell}}=\mu/\omega$. For tidal forcing at Europa’s orbital frequency $\omega = 2.0\times 10^{-5}$~rad~s$^{-1}$ and taking elastic shear modulus $\mu = 3.3$~GPa, maximum dissipation occurs at viscosity $\eta_{\rm{Maxwell}}= 1.6\times 10^{14}$~Pa~s. For Enceladus, a higher orbital frequency $\omega=5.3\times 10^{-5}$~rad~s$^{-1}$ leads to maximum dissipation at a lower viscosity $\eta_{\rm{Maxwell}} = 6.2\times 10^{13}$~Pa~s. As discussed in Section \ref{sec. rheology}, the viscosity at the base of the shell $\eta_{\rm{melt}}$ is taken as a range from $10^{13}$ to $10^{15}$~Pa~s covering a range of grain sizes. Since $\eta_{\rm{Maxwell}}$ is close to $\eta_{\rm{melt}}$, we expect that our tidal heating model will produce maximum tidal dissipation near the base of the shell. In summary, the amount of tidal heating and the shell temperature profile depends strongly on the assumed melt viscosity. 

\subsection{Temperature structure}
\label{sec. temperature}
Conductive heat transport through a tidally heated shell defines its temperature profile, which controls its viscosity profile. Since tidal heating depends on the viscosity profile, we need to solve for tidal dissipation and temperature profile simultaneously. In a steady state, assuming that heat transport is dominated in the radial direction, the temperature and tidal dissipation satisfy the 1D conductive heat equation in spherical coordinates:

\begin{equation}
    \frac{1}{r^2} \frac{\partial}{\partial r} \left(r^2 \kappa(T) \frac{\partial T}{\partial r} \right) + h_{\rm{tide}}(r) = 0
    \label{eq. heat equation},
\end{equation}

\noindent where the source term $h_{\rm{tide}}(r)$ is the surface averaged volumetric tidal dissipation rate found in Section \ref{sec. tidal} and thermal conductivity for pure water ice is given by $\kappa(T) = 0.4685 + 488.12/T$ in W~$\mathrm{m}^{-1}$~$\mathrm{K}^{-1}$ \citep{hobbs2010ice}. 

Once the tidal dissipation term $h_{\rm{tide}}(r)$ is known, Eq. \ref{eq. heat equation} can be solved numerically as a boundary value problem with temperatures set at the surface and base of the shell. The temperature at the base is assumed to be $T_b = 273$~K and the surface temperature $T_s$ is calculated as an equilibrium temperature using a fast rotator approximation, which gives $T_s = 92$~K for Europa and $T_s = 59$~K for Enceladus (see Appendix \ref{apdx surface temperature}). For the range of viscosity values $\eta_{\rm{melt}}$ considered, \cite{tobie2003tidally} and \cite{roberts2008tidal} show for Europa and Enceladus, respectively, that the tidal heating rate at the base is larger than the heat flux from radiogenic heating in the silicate mantle. However, the tidal heat production in the rocky mantles can be significant if the rocky mantle is loosely consolidated \citep{roberts2015fluffy}. In our modeling, however, we neglect mantle dissipation and the heat flux from the ocean for simplicity, considering only the tidal heat produced within the icy shell. Thus, the mantle is assumed to have an infinite viscosity and, thus, deforms only elastically. 

To solve for a steady-state solution of the heat conduction equation, we converge the tidal dissipation and temperature profiles with the following iterative approach:

\begin{enumerate}
\item The solution to the boundary value problem of Eq. \ref{eq. heat equation} excluding the tidal heating term ($h_{\rm{tide}}(r)=0$) is set as an initial guess for the temperature profile.
\item The temperature profile obtained is used to find the viscosity profile with Eq. \ref{eq. viscosity}, and tidal dissipation $h_{\rm{tide}}(r)$ is computed.
\item Then, the full Eq. \ref{eq. heat equation} is used with $h_{\rm{tide}}(r)$ and a new temperature profile is found as the solution to the boundary value problem. 
\end{enumerate}

Steps 2 and 3 are repeated until the temperature profile has converged or the temperature at any point within the shell has exceeded 273~K. Temperatures above the melting point in portions of the shell would cause different flow mechanisms to dominate due to partial melting, which would invalidate our ice rheology assumption (Eq. \ref{eq. viscosity}). Thus, we exclude such temperature profiles from the relaxation modeling. 

\subsection{Viscous relaxation model}
\label{sec. relaxation}

Given a shell structure, we model the viscous relaxation of the icy shell as an incompressible fluid in a spherical shell with self-gravitation following the methodology of \cite{hager1989constraints}. The icy shell is described by a sequence of layers of constant density and viscosity as prescribed by the viscosity profile. We have tested how the relaxation study results depend on the number of layers and found that 80 layers are needed to arrive at a converged solution. Incompressible Stokes flow along with Poisson's equation for gravitational potential are formulated as a system of six linear differential equations for a spherically symmetric shell \citep{hager1989constraints}. The state vector is given as radial functions of radial and poloidal velocities, radial normal stress, and poloidal shear stress as coefficients of vector spherical harmonics. Gravitational potential perturbation and its derivative are given as coefficients of scalar spherical harmonics. Given boundary conditions at the surface and base of the shell, the system of equations can be solved with a propagator matrix method \citep{gantmacher2005applications}. Thus, our layered model can be expressed as a single linear system that relates the state vector at the ocean-ice interface to the state vector at the ice-outer space interface. A full description of the viscous relaxation model and the implementation of boundary conditions (Section \ref{sec. BC}) are given in Appendix \ref{apdx stokes flow}.

\subsubsection{Boundary conditions}
\label{sec. BC}

\begin{figure}
    \centering
    \includegraphics[width=\columnwidth]{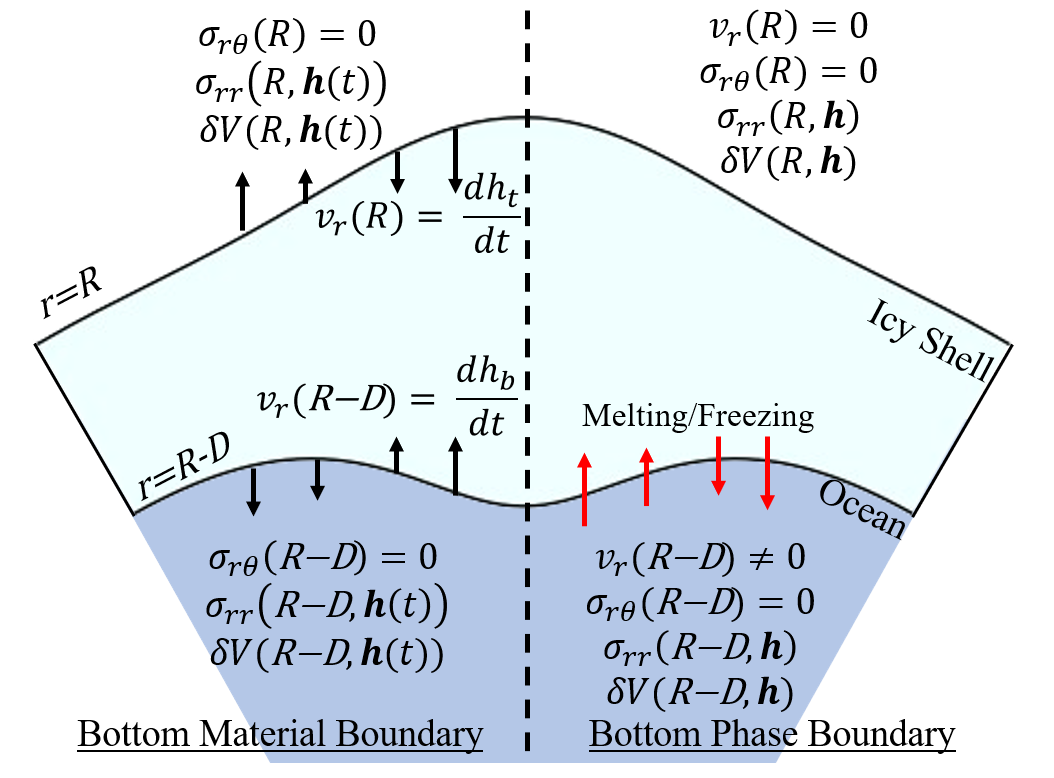}
    \caption{Illustration of bottom material (left) and bottom phase (right) boundary conditions. The icy shell is shown with outer radius $R$ and thickness $D$ with some hypothetical topography (not to scale). $\mathbf{h} = [h_t, h_b]^T$ are the shapes of the surface and base of the icy shell, $v_r$ is radial velocity, and $\sigma_{rr},\sigma_{r\theta}$ are radial stress and poloidal shear stress perturbations as coefficients of expanded vector spherical harmonics. $\delta V$ is gravitational potential perturbation as a coefficient of expanded scalar spherical harmonics. Note that boundary conditions for bottom material boundary are functions of time, while bottom phase boundary is in a dynamic equilibrium.}
    \label{fig. bc}
\end{figure}

The two boundaries of the icy shell are the ice-outer space interface at the surface of the moon and the ice-ocean interface at the base of the shell.  Under the formulation of \cite{hager1989constraints}, the shapes of the interfaces are represented as loads applied at the artificial spherical boundaries $r=R$ and $r=R-D$, for the mean radius $R$ and shell thickness $D$. The amplitude of this topography is assumed to be small compared to its wavelength $2\pi R/\sqrt{n(n+1)} \approx 2\pi R/n$ for a spherical harmonic degree $n$. The loads generated by topography create perturbations in radial stress and gravity potential. The surface ice-outer space boundary is a material boundary. Thus, the radial velocity at the boundary directly affects the surface topography. Finally, free-slip boundary conditions are imposed at both boundaries, setting shear stress to zero.

At the bottom ice-ocean boundary, two types of boundary conditions are considered as illustrated in Fig.~\ref{fig. bc}. A bottom material boundary condition, which we will refer to as ``bottom material boundary'', assumes that no material crosses the water-ice interface through phase transitions. A bottom material boundary is similar to the boundary condition described by \cite{hager1979kinematic} for an isostatic rebound problem, and ``constant load'' described by \cite{beuthe2020isostasy}. With this material condition, the radial component of the flow at the base of the icy shell will create or remove topography. The two interfaces will approach hydrostatic equilibrium over time. On the other hand, a bottom phase boundary condition, or ``bottom phase boundary'', allows for a flow across the ice-ocean interface. Under this boundary condition, described by \cite{vcadek2019airy} to model the shell of Enceladus, the surface non-hydrostatic topography is maintained in a dynamic equilibrium by the flow at the base of the shell that is balanced by phase transitions (melting or freezing):

\begin{equation}
    \hat{\mathbf{n}}\cdot \mathbf{v} L \rho_1 = \hat{\mathbf{n}}\cdot (\mathbf{q}_1-\mathbf{q}_2),
    \label{eq. phase boundary}
\end{equation}

\noindent where $\hat{\mathbf{n}}$ is a unit vector normal to the interface, $\mathbf{v}$ is the velocity of ice flow, $L$ is the latent heat of fusion for ice, $\rho_1$ is the density of the shell, and $\mathbf{q}_1,\mathbf{q}_2$ are heat fluxes from the icy shell and ocean respectively. Our bottom phase boundary is the same as ``constant shape'' described by \cite{beuthe2020isostasy}. 

We solve for the Stokes flow using each of these boundary conditions with the goal of obtaining the asymptotic shape of the icy shell at spherical harmonic degree $n$ for the calculation of gravity-topography admittance (Section \ref{sec. admittance}). The shape of the icy shell at the surface and base are expanded in spherical harmonics at radii $R$ and $R-D$, and their spherical harmonic coefficients $h_t$ and $h_b$ are referred to as shape coefficients or amplitudes, where the dependence on $n$ is implied. We note that the term ``shape'', when used in describing our computations, is distinct from ``topography'', which refers to the elevation difference between shape and the equipotential surface. However, we use the term ``topography'' liberally to describe ``shape'' throughout the paper when describing features at the boundaries or topography support mechanisms. The shape of the ice-ocean boundary is described by the shape ratio $w_n = -h_b/h_t$. This is different from the topographic ratio used in \cite{vcadek2019airy}, which describes the ratio of topographies. The negative sign in the definition of $w_n$ is a matter of convention and we use it to make $w_n$ positive for the conventional Airy isostasy case. 

Solving for the Stokes flow under the described boundary conditions requires two different approaches. For the bottom material boundary, the shapes of the interfaces are functions of time and eventually relax to hydrostatic equilibrium. Applying the bottom material boundary, the propagator matrix solution of the Stokes flow equations \citep{hager1989constraints} can be rearranged as a $2\times 2$ system of ordinary differential equations for radial velocities $\frac{d h_t}{d t}$ and $\frac{d h_b}{d t}$ (see Appendix \ref{apdx stokes flow} for details). The solution for the shape amplitudes $\mathbf{h}(t) = [h_t(t), h_b(t)]^T$ is given as a sum of two decaying exponentials describing the time evolution of the interfaces: 

\begin{equation}
    \mathbf{h}(t) = \mathbf{A} e^{-\gamma_1 t} + \mathbf{B} e^{-\gamma_2 t},
    \label{eq. shell evolution}
\end{equation}

\noindent where $\mathbf{A}=[A_s,A_b]^T$, $ \mathbf{B}=[B_s,B_b]^T$ are the eigenvectors of the system, and $\tau_1 = 1/\gamma_1, \tau_2=1/\gamma_2$ are the characteristic decay times. We set $\tau_1<\tau_2$, so that the second term in Eq.~\ref{eq. shell evolution} corresponds to the longer decay time. Two modes arise from Eq. \ref{eq. shell evolution}, which are described by \cite{hager1979kinematic} as symmetric and antisymmetric modes of relaxation. In the symmetric mode, the two interfaces move in the same direction. For example, if a mountain were placed on the surface and allowed to relax, a crustal root would initially form at the base of the icy shell. Conversely, in the antisymmetric mode, the two interfaces move in opposite directions. In this mode of relaxation, the height of the mountain and the crustal root will decrease with time, approaching hydrostatic equilibrium. We obtain the asymptotic state of the shell from the eigenvector $[B_s,B_b]^T$ of Eq.~\ref{eq. shell evolution} corresponding to the longer decay time $\tau_2$. The degree-dependent shape ratio is given by $w_n = -B_b/B_s$.

For the bottom phase boundary, there is no time-dependent solution of the interfaces. Instead, we solve for $h_t$ and $h_b$ as constants for the dynamic equilibrium between ice flow and phase transitions at the base. In this state, the radial velocity of the surface interface is set to zero to maintain the shape of the surface. However, the radial velocity at the base can be nonzero as long as the flow across the bottom interface is balanced by phase transitions satisfying Eq.~\ref{eq. phase boundary}. By setting surface shape $h_t$ to unity, the propagator matrix solution to the system of \cite{hager1989constraints} can be rearranged into a system of four equations with four unknowns: bottom shape $h_b$, bottom radial velocity, bottom poloidal velocity, and surface poloidal velocity. Thus, the shape ratio is equal to $w_n = -h_b$. We note that with an observed value of $h_t$, assuming that the shell is in this dynamic equilibrium, we can find bottom radial velocity, and thus, heat flux from Eq. \ref{eq. phase boundary}. Our method of applying the bottom phase boundary is similar to the spectral method used by \cite{vcadek2019airy}. 

\subsection{Gravity-topography admittance}
\label{sec. admittance}
Finally, we compute the gravity-topography admittance using the shape amplitudes of viscously relaxed icy shells. We follow an approach of \cite{ermakov2017constraints} used to model the admittance of Ceres. Assuming shape amplitude is small relative to its wavelength, we use a mass-sheet approximation to get a linear relationship between gravity and shape. Summing the contributions to gravity from the shape of the surface and bottom boundaries of the shell, and dividing by surface shape, we find an expression for admittance $Z_n$ at degree $n$ for shape ratio $w_n$:

\begin{equation}
    Z_n=\frac{GM}{R^3}\cdot\frac{3(n+1)}{(2n+1)}\cdot\left[\frac{\rho_{1}}{\bar{\rho}}-\frac{\Delta\rho}{\bar{\rho}}\left(\frac{R-D}{R}\right)^{n+2}w_n\right],
    \label{eq. Admittance}
\end{equation}

\noindent where $\bar{\rho}$ is the mean density of the moon, $\rho_1$ is the density of the icy shell, $\rho_2$ is the ocean density, and $\Delta\rho = \rho_2 - \rho_1$ is the ice-ocean density contrast. See Appendix \ref{apdx admittance} for derivation. Thus, with a degree-dependent shape ratio $w_n$ obtained from the viscous relaxation model, we find the admittance spectrum for a given shell structure.

We have two points of comparison for admittance. First, in the classical case of Airy isostatic compensation and assuming Cartesian geometry, the amplitude of the bottom shape is related to the amplitude at the surface by the ratio of the shell density to the density contrast, leading to $w_n = \rho_1/\Delta\rho$. Airy compensated values of admittance can inform us of topography supported by buoyancy forces acting on the bottom topography, or isostatic roots. The second limiting case is that of uncompensated surface topography, for which there is no corresponding bottom topography $h_b=0$. Therefore, $w_n = 0$ for uncompensated surface topography. The admittance for the uncompensated topography provides an upper bound for admittance for a given density of a shell. Uncompensated values can indicate that topography is supported by stresses within the shell. We note, however, that Airy admittance does not necessarily provide a lower bound on admittance. 

\section{Results}

\subsection{The effect of the bottom boundary condition}

\begin{figure}[h!]
    \centering
    \includegraphics[width=\columnwidth]{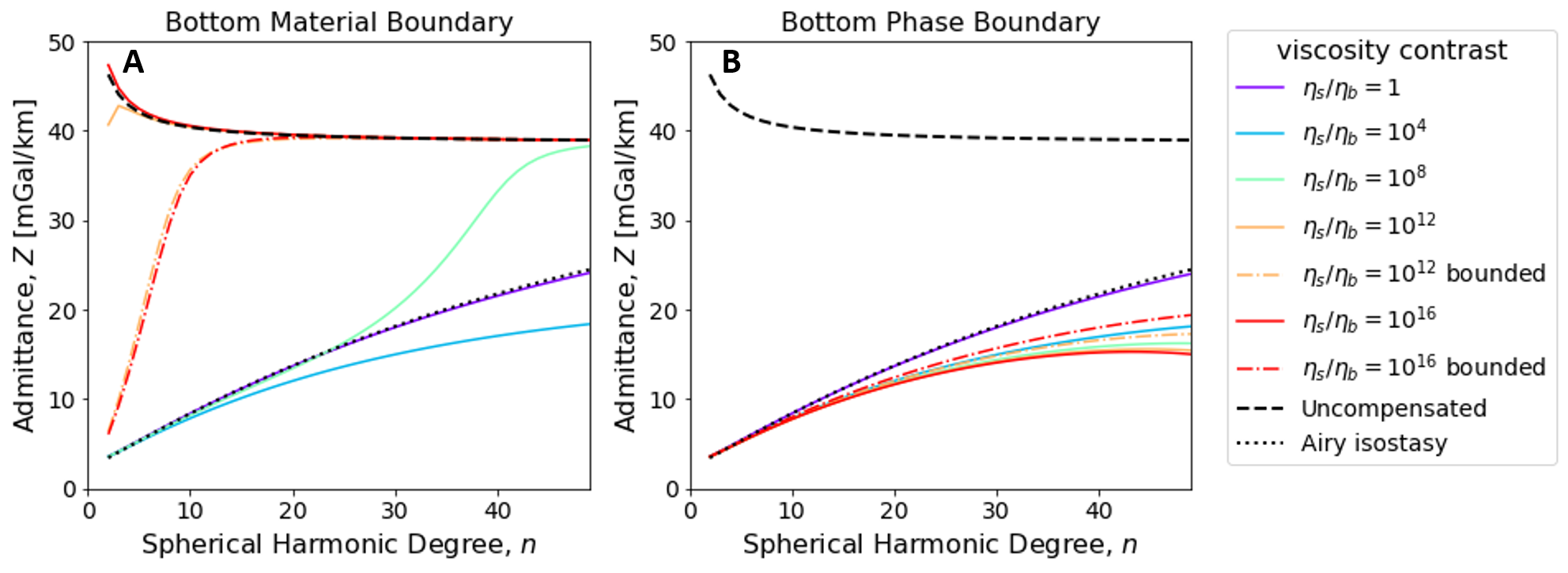}
    \caption{Admittance spectra of Europa with a 30~km icy shell. We show admittance spectra with both ({\bf{A}}) bottom material and ({\bf{B}}) bottom phase boundary conditions for a range of viscosity gradients. Viscosity is assumed to decrease exponentially with depth, with the viscosity contrast from the surface to base $\eta_s/\eta_b$ where $\eta_b = 10^{14}$~Pa~s is fixed. For the two high viscosity contrasts, admittance spectra for bounded viscosity $\eta_{\rm{bound}} = 10^{22}$~Pa~s is shown. The uniform viscosity case ($\eta_s/\eta_b = 1$) is plotted in purple. Admittance spectra assuming uncompensated topography and Airy isostasy are shown by the black dashed and dotted curves, respectively. Admittance values approach uncompensated values at lower spherical harmonic degrees with steepening viscosity gradient.}
    \label{fig.adm viscosity}
\end{figure}

We begin by studying the effects of bottom material boundary and bottom phase boundary on admittance. A model of Europa with a 30-km shell is created for a range of viscosity profiles. Viscosity is assumed to decrease exponentially with depth. We vary the viscosity contrast between the top and the bottom of the shell ($\eta_s/\eta_b$). Fig.~\ref{fig.adm viscosity} compares admittance spectra for both boundary conditions. We observe that for the uniform viscosity case ($\eta_s/\eta_b=1$) shown in purple, the two boundary conditions produce admittance spectra that are nearly identical to the Airy isostasy case shown as the dotted curve. However, as the viscosity gradient steepens, the two boundary conditions start to diverge at higher spherical harmonic degrees. 

The bottom phase boundary produces lower admittance values than for bottom material boundary. Lower admittance values correspond to larger values of shape ration $w_n$. Thus, in this case, the shell has larger amplitude bottom topography relative to its surface topography compared to the bottom material boundary case. This is expected as the dynamic equilibrium between phase transitions and ice flow sustains more bottom topography compared to a material boundary. However, large amplitude topography at the surface could indicate that an unrealistic heat flux is needed to sustain melting and freezing. Bottom phase boundary can be justified if a sufficient heat source is present at the appropriate wavelength and magnitude to support the observed surface topography. This would be difficult to reconcile with tidal heating. The tidal potential is dominated by the degree~2 component. The higher-degree terms decrease quickly in amplitude. For example, for Europa, the degree~3 term is 1/430 the degree~2 term \citep{sabadini2016global}. Therefore, tidal heating could drive phase transitions only at low degrees, unless there are significant heterogeneity within the shell or in the heat flow from the ocean. 

\subsection{The effect of the viscosity gradient within the shell}

\begin{figure}
    \centering
    \includegraphics[width=\columnwidth]{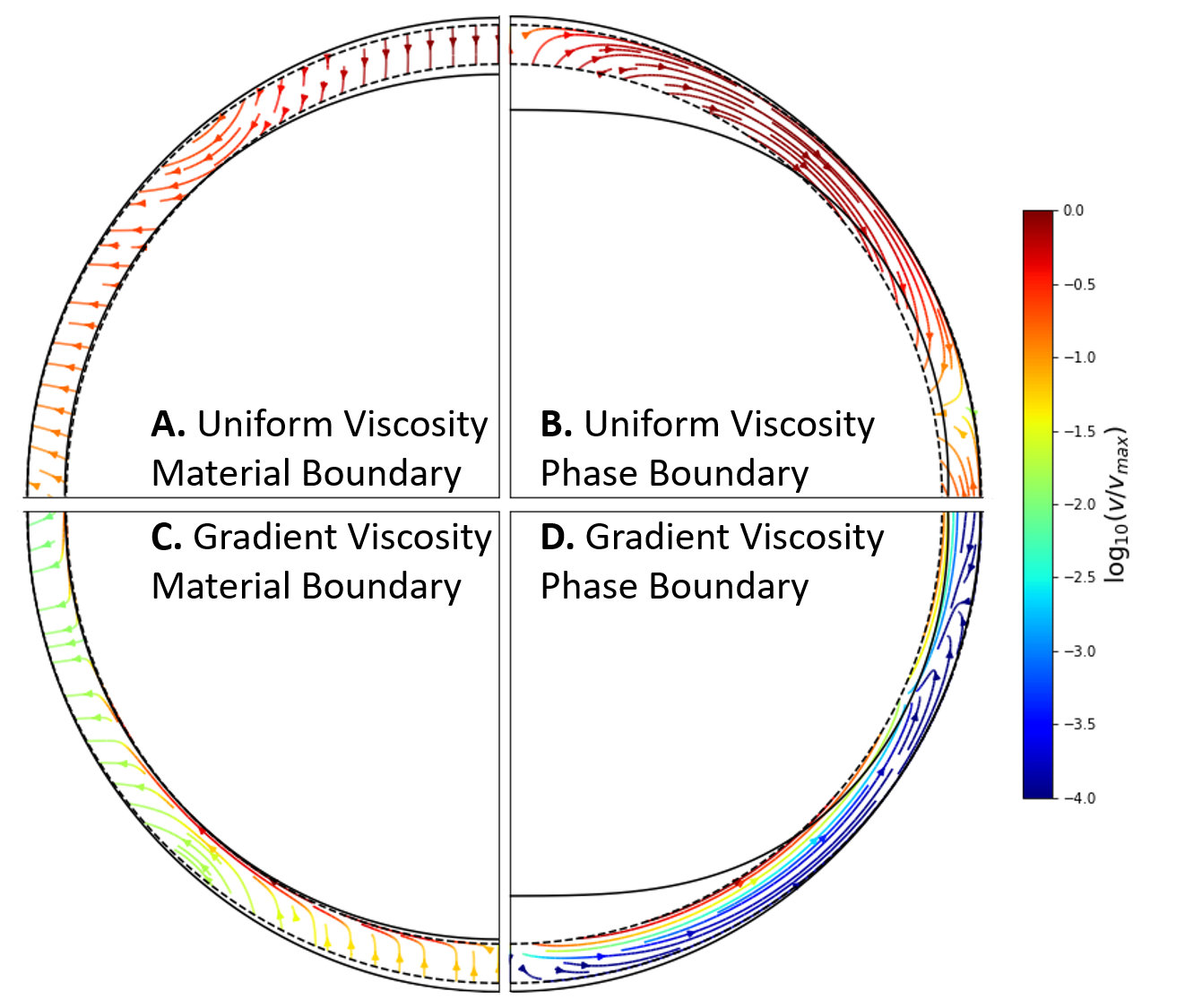}
    \caption{Snapshot of the flow within a 21-km shell of Enceladus driven by the viscous relaxation of an artificial mountain. A large axially symmetric mountain, expressed as a Gaussian ($R+h e^{-2.5\theta^2}$ for height $h$, colatitude $\theta$, and full width at half maximum of $\approx 60$ degrees), was placed at the pole expanded in spherical harmonics. It was then relaxed under bottom material ({\bf{A}},~{\bf{C}}) and bottom phase ({\bf{B}},~{\bf{D}}) boundary conditions for a uniform viscosity shell ({\bf{A}},~{\bf{B}}) and gradient viscosity shell ({\bf{C}},~{\bf{D}}). The gradient viscosity shell has a viscosity contrast of $\eta_s/\eta_b = 10^{8}$~Pa~s with $\eta_s=10^{22}$~Pa~s. The streamlines represent the direction and colors represent the magnitude of velocity of the ice flow found from our Stokes flow model. For viscous relaxation under bottom material boundary, the snapshot was taken when the dominant motion in the shell is downwards (symmetric mode of relaxation). For bottom phase boundary, the figure shows the state of dynamic equilibrium.}
    \label{fig.flowfield}
\end{figure}

Here, we explore the effect of the shell viscosity structure on admittance using exponential viscosity profiles of variable steepness. To illustrate the effect of the viscosity profile, Fig.~\ref{fig.flowfield} shows the velocity field comparing the ice flow within a uniform and gradient-viscosity shell for the bottom material and phase boundaries. The flow within the icy shell of Enceladus is shown for illustrative purposes, as the Enceladus' shell is thicker compared to its radius relative to Europa's shell. The flow pattern for the four shown cases is qualitatively similar for Europa.

For uniform-viscosity shells, the flow is quasi-uniform in amplitude throughout the shell beneath the topographic load. (Fig.~\ref{fig.flowfield}A and B). On the other hand, a viscosity gradient in the shell causes a strong lateral flow concentrated at the low-viscosity base for both types of bottom boundary conditions (Fig.~\ref{fig.flowfield}D and C). For bottom material boundary, this lateral flow rapidly relaxes basal topography compared to the relaxation of the high-viscosity near-surface layers. For bottom phase boundary, there is a net freezing at the base of the shell associated with the topographic load at the pole. This freezing is balanced by upward radial flow such that the bottom interface remains stationary. 

We observe that as the viscosity gradient steepens, admittance values approach high, uncompensated values at progressively lower degrees for bottom material boundary (Fig.~\ref{fig.adm viscosity}A). Uncompensated values indicate that there is little basal topography relative to surface topography. Thus, surface topography is not supported by buoyancy as in the Airy isostasy case. Instead, viscous stresses within the shell provide the dominant topography support mechanism. For very large viscosity contrasts (the red bounded and unbounded curves for the bottom material boundary, Fig.~\ref{fig.adm viscosity}A), the high relative viscosity at the surface impedes the relaxation of surface topography even at large scales while the bottom topography relaxes, leading to uncompensated admittance values at low degrees. For the bottom phase boundary, we find that admittance spectra are not very sensitive to the steepness of the viscosity gradient (Fig.~\ref{fig.adm viscosity}B), especially at the lowest degrees. This observation also holds for admittance spectra of Enceladus, and is consistent with the observation of similar shape amplitudes of the bottom interfaces seen in Fig.~\ref{fig.flowfield}B and D. 

We note that admittance spectra are only affected by the viscosity contrast $\eta_s/\eta_b$ and not the actual values of $\eta_s$ or $\eta_b$, which, instead, affect the timescales of relaxation, with higher viscosities increasing the time it takes to reach the asymptotic state of topography. The choice of the top viscosity bound affects the admittance spectra, with higher values for $\eta_{\rm{bound}}$ increasing the viscosity contrast, causing admittance to approach uncompensated values at lower degrees.

\subsection{Tidal heating influence on viscosity}  

\begin{figure}
    \centering
    \includegraphics[width=\columnwidth]{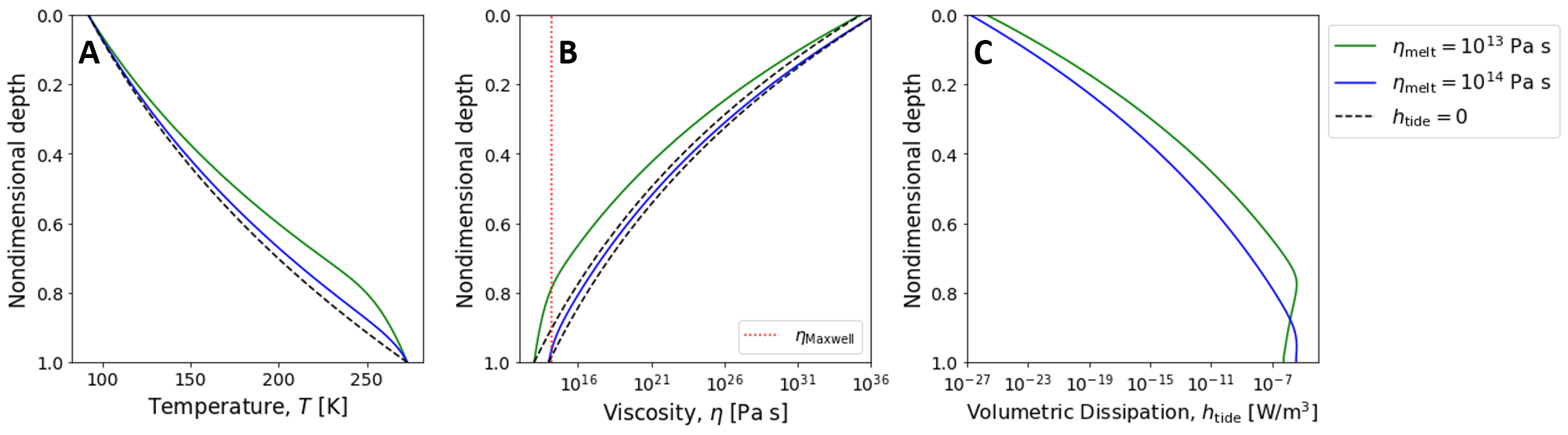}
    \caption{Effects of tidal heating on temperature and viscosity in a 30-km shell of Europa. {\bf{A:}} Conductive temperature profiles converged from iterative process for melting point viscosity $\eta_{\rm{melt}} = \{10^{13},10^{14}\}$~Pa~s and for a case without tidal heating ($h_{\rm{tide}} = 0$). {\bf{B:}} Unbounded temperature-dependent viscosity profiles calculated from the temperature profiles of {\bf{A}} with Eq. \ref{eq. viscosity}. The viscosity at which Maxwell time is crossed is shown as the red dotted line. The case without tidal heating is shown for both choices of $\eta_{\rm{melt}}$. {\bf{C:}} Surface averaged volumetric tidal dissipation rate is shown for $\eta_{\rm{melt}} = \{10^{13}$, $10^{14}\}$~Pa~s. Maximum tidal dissipation can be seen occurring at the depth where the viscosity profile crosses $\eta_{\rm{Maxwell}}$}
    \label{fig.tidal heating}
\end{figure}

We proceed from simple viscosity gradients to more realistic viscosity profiles by including a temperature-dependent viscosity and tidal heating. We incorporate the effect of tidal heating on a conductive temperature profile of the icy shell using a steady-state solution obtained from the iterative method described in Section \ref{sec. temperature}. Fig.~\ref{fig.tidal heating} shows the effect of including tidal heating on the temperature (Fig.~\ref{fig.tidal heating}A), viscosity (Fig.~\ref{fig.tidal heating}B), and volumetric tidal dissipation rate (Fig.~\ref{fig.tidal heating}C) for two choices of $\eta_{\rm{melt}}$. As mentioned in Section \ref{sec. tidal}, tidal dissipation is maximized if the forcing period is near the Maxwell time of the material being deformed. The viscosity for which this occurs for Europa ($\eta_{\rm{Maxwell}} \approx 1.6 \times 10^{14}$~Pa~s) is shown as the dotted line in Fig.~\ref{fig.tidal heating}B. The viscosity profiles of the 30-km shells modeled with two values of $\eta_{\rm{melt}}$ cross $\eta_{\rm{Maxwell}}$ near the base of the shell. For $\eta_{\rm{melt}} = 10^{13}$~Pa~s, this intersection occurs at $\approx$78\% of the shell's depth, which corresponds to a maximum in tidal dissipation as expected. For the case with $\eta_{\rm{melt}}=10^{14}$~Pa~s, maximum dissipation occurs closer to the base of the shell. This causes tidal dissipation to be lower throughout the shell, which is seen in the temperature profile in Fig.~\ref{fig.tidal heating}A, where $\eta_{\rm{melt}}=10^{14}$~Pa~s causes the temperature to be closer to the no tidal heating case. Although not shown in Fig.~\ref{fig.tidal heating}, $\eta_{\rm{melt}}=10^{15}$~Pa~s further decreases the influence of tidal heating, and the temperature profile becomes nearly identical to the case without tidal heating. In summary, the choice of $\eta_{\rm{melt}}$, and thus the assumption of the grain size of ice, has a large effect on tidal heating and, therefore, the temperature structure of the shell.

\subsection{Shell thickness}
 
\begin{figure}
    \centering
    \includegraphics[width=\columnwidth]{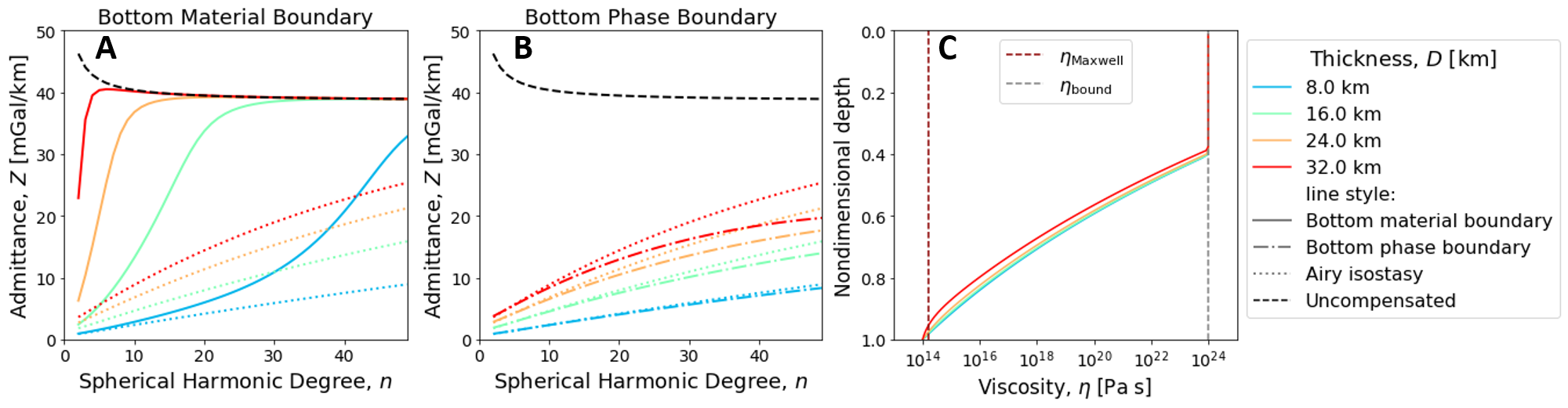}
    \caption{Admittance spectra and viscosity profiles for tidally heated shells of Europa with different shell thicknesses. {\bf{A:}} Admittance for bottom material boundary (solid curve) and Airy isostasy (dotted curve) are shown for different shell thicknesses with solid lines. Admittance spectra assuming uncompensated topography are shown with a dashed curve. {\bf{B:}} Admittance spectra for bottom phase boundary (dot-dashed curve) are shown for different shell thicknesses. {\bf{C:}} Viscosity profiles are plotted against depth normalized to shell thickness, $D$. A value of $\eta_{\rm{melt}}=10^{14}$~Pa~s is chosen and a viscosity bound of $\eta_{\rm{bound}} = 10^{24}$~Pa~s is imposed. }
    \label{fig.adm thickness}
\end{figure}

Finally, we use our complete model with temperature-dependent viscosity and tidal dissipation to study the sensitivity of admittance to the shell thickness. Predictions of Europa’s shell thickness vary from local estimates of several km to more than $30$~km as a global average \citep{billings2005great}. We consider mean thicknesses between $D =$ 8 and 35 km. Depending on the viscosity at the base $\eta_{\rm{melt}}$, the amount of tidal heating for thick shells causes temperature profiles to not converge with our iterative method described in Section \ref{sec. temperature}. For $\eta_{\rm{melt}}=10^{14}$~Pa~s, a temperature profile did not converge above a shell thickness of 32~km due to temperatures exceeding the melting point in our iterative method.

The shell thickness has a strong effect on the admittance spectrum, which is illustrated in  Fig.~\ref{fig.adm thickness}. The bottom material boundary condition is used in Fig.~\ref{fig.adm thickness}A and the bottom phase boundary condition is used in Fig.~\ref{fig.adm thickness}B. We observed that increasing shell thickness lowers viscosity at relative depths (Fig.~\ref{fig.adm thickness}C) as thicker shells generate more heat from tidal dissipation. 

To understand the behavior of admittance and topography support mechanisms at different wavelengths, we apply Jeffreys' theorem. It states that the minimum stress difference required to support a surface load is $\approx 1/3$ of the load and is concentrated in a region with dimensions comparable to the width of the load \citep{melosh2011planetary}. Thus, a surface topographic load samples the shell beneath it to a depth comparable to its wavelength. A short-wavelength load feels only the near-surface viscosity value. A load with wavelength comparable to the shell thickness samples deeper into the shell and is sensitive to its viscosity profile. For a load with a wavelength much longer than the shell thickness, the shell acts essentially as a membrane. Thus, such a load is not sensitive to the viscosity profile. 

As shell thickness increases, admittance for bottom material boundary (Fig.~\ref{fig.adm thickness}A) approaches uncompensated values at progressively lower degrees or longer wavelengths. For thicker shells, it is difficult to propagate buoyancy-produced stresses from the base of the shell to support surface topography, causing topography at progressively longer wavelengths to be supported by stresses within the shell. Additionally, wavelengths much longer than the thickness of the shell do not sense the viscosity variation throughout the shell, thus yielding an Airy-like admittance. This is seen for the 8-km shell at low degrees ($n<15$) and at the lowest degrees for a 16-km shell (Fig.~\ref{fig.adm thickness}A). For shorter wavelengths, admittance becomes sensitive to the steep viscosity profile resulting from a tidally heated conductive shell and approaches uncompensated values. 

For bottom phase boundary (Fig.~\ref{fig.adm thickness}B), admittances are generally closer to Airy isostasy admittance values. As the shell thickness increases, the degree at which admittance deviates from the Airy isostasy becomes progressively lower, however at low degrees ($n<15$), values of admittance are similar to that of Airy isostasy. For thicker shells, the divergence between boundary conditions occurs at lower degrees. In a scenario where the shell is thick, measurements showing a dip in admittance at a high degree---characteristic of bottom phase boundary---could indicate the presence of a heat source pattern sufficient to maintain the dynamic equilibrium of bottom phase boundary. 

\section{Discussion}
\subsection{Topography support mechanism}

Admittance can provide insight about topography support mechanism. High, uncompensated values of admittance can indicate that surface topography that is supported by viscous stresses within the shell. Low admittance could indicate that topography is supported by buoyancy, as predicted from the Airy isostasy model. Alternatively, low admittance could indicate that surface topography is supported by flow arising from phase transitions at the base of the shell. In this case, the heat flux pattern controls the rate of basal melting and freezing. Further study on the distribution of basal heat flux is needed to understand the wavelengths at which bottom phase boundary could be applicable. If tidal heating in the shell is small, for example due to high values of $\eta_{\rm{melt}}$, the influence of radiogenic and tidal heating from the mantle and ocean \citep{roberts2015fluffy,rekier2019internal,rovira2019tidally} may become more important. Ocean circulation patterns resulting from mantle heat sources and salinity gradients could affect the distribution of heat at the base of the shell \citep{kang2020differing,kang2021does}. 


\subsection{Ice rheology}

Temperature-dependent viscosity for a tidally heated conductive shell yields a steep viscosity gradient within the shell. The depth at which Maxwell time is crossed influences the distribution of tidal dissipation, which, in turn, determines the extent of the high-temperature, low-viscosity region at the base of the shell. On Europa, tidal dissipation is concentrated at the base of the shell, where viscosity is low and is close to $\eta_{\rm{Maxwell}}$. We find that the low viscosity at the base causes rapid relaxation of bottom topography by lateral flow, leading to uncompensated surface topography. Thus, the Airy isostasy model is likely not suitable for interpreting admittance except at the longest wavelengths, which are not sensitive to the steep viscosity gradient, or where the bottom phase boundary is applicable.

Ice deformation mechanisms are difficult to study in the laboratory, and different non-Newtonian flow mechanisms such as grain boundary sliding may be more appropriate when considering solid-state convection \citep{barr2009heat}. Although our analytical viscous relaxation model does not allow for convection, we expect that convection at the base of the shell would effectively relax shorter wavelength bottom topography resulting in uncompensated values of admittance.

In addition to the flow mechanism, the grain size of ice has a large effect on the viscosity profile, tidal dissipation, and thus the temperature structure of the shell. The viscosity of ice at the base of the shell at the melting temperature depends on this poorly constrained grain size of ice. For Europa and Enceladus, the viscosity at which Maxwell time is crossed, $\eta_{\rm{Maxwell}}$, lies within the range of melt viscosities $\eta_{\rm{melt}} = 10^{13}-10^{15}$~Pa~s typically considered. We find that tidal heating has little influence on temperature structure when $\eta_{\rm{melt}}>\eta_{\rm{Maxwell}}$. If $\eta_{\rm{melt}}<\eta_{\rm{Maxwell}}$, the influence of $\eta_{\rm{melt}}$ on temperature profile is amplified depending on the extent of the low viscosity region and the relative depth at which $\eta_{\rm{Maxwell}}$ is crossed. The melting temperature $T_{\rm{melt}}$ changes with pressure and affects the value of $\eta_{\rm{melt}}$ and the temperature gradient. However, changes in grain size likely have a larger effect on $\eta_{\rm{melt}}$, thus we can approximate different melting temperatures with the range of $\eta_{\rm{melt}}$ considered.

The diffusion creep temperature-dependent viscosity predicts high viscosities at the surfaces of Europa and Enceladus, implying very slow relaxation of surface topography leading to uncompensated values of admittance. Properties of shallow sub-surface ice such as fracturing, porosity, and impurities affect deformation mechanisms, viscosity, and thermal conductivity. An insulating layer, considered by \cite{bland2012enceladus} in their study of crater relaxation on Enceladus, could raise the shallow sub-surface temperature and, thus, lower viscosities near the surface. In addition, isolated periods of warming may have occurred in the past \citep{bland2012enceladus}. Such warming episodes could have facilitated viscous relaxation affecting admittance. Thus, measuring admittance and comparing the inferred viscosity profile to predictions from the crater relaxation study could validate the notion of past heating episodes on Enceladus.

\subsection{Comparing Europa to Enceladus}
\label{sec. compare}

\begin{figure}
    \centering
    \includegraphics[width=\columnwidth]{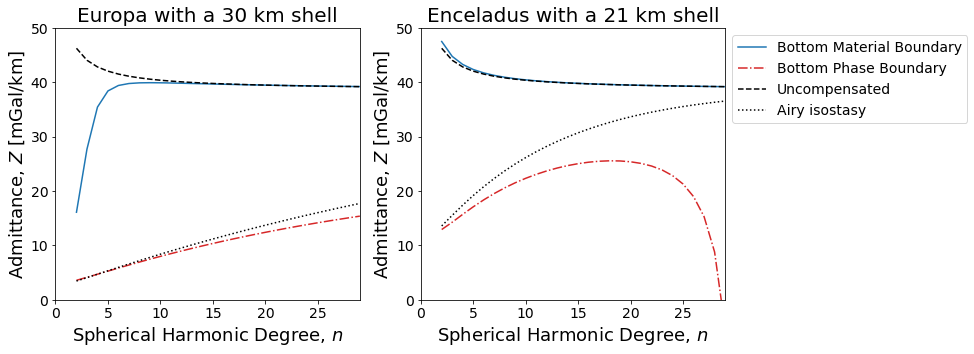}
    \caption{Comparison of admittance spectra for models of Europa and Enceladus. Shell thicknesses of 30~km for Europa (left) and 21~km for Enceladus (right) are assumed. Admittance spectra for bottom material (solid) and bottom phase boundaries (dot-dash) are shown. Admittance assuming for the uncompensated (dash) and Airy isostasy (dot) cases are also shown. Melt viscosity $\eta_{\rm{melt}} = 10^{14}$~Pa~s and viscosity bound $\eta_{\rm{bound}}=10^{24}$~Pa~s are used. The effect of Enceladus' thicker shell relative to Europa is seen in the admittance spectra for the bottom material boundary, where the transition from low to uncompensated values occurs at lower degrees.}
    \label{fig.adm compare}
\end{figure}

We now apply our methods to a model of Enceladus and compare it to Europa. We assume a shell thickness of 21~km for Enceladus taking the central value from \cite{hemingway2019enceladus}, and compare it to Europa with a 30~km shell using melt viscosity $\eta_{\rm{melt}}=10^{14}$~Pa~s for both ocean worlds. Enceladus has a thicker shell in proportion to its radius compared to Europa. We show a comparison of admittance spectra in Fig.~\ref{fig.adm compare}. Since the wavelength is inversely proportional to degree, the same degree on Enceladus corresponds to a shorter wavelength than on Europa. Thus, we expect admittance spectra of bottom material boundary for Enceladus to approach uncompensated values at a lower spherical harmonic degree than for Europa due to shorter wavelengths at low degrees for Enceladus sampling the viscosity gradient of the shell. Comparing admittance of bottom phase boundary, we see that admittance for Enceladus deviates from the Airy isostasy case at low degrees more than for Europa, perhaps making it easier to distinguish topography support mechanism from future observations.

For tidal heating, the key differences between the two moons are orbital frequency and surface temperature. The higher orbital frequency of Enceladus causes $\eta_{\rm{Maxwell}} = 6.2\times 10^{13}$~Pa~s to be lower than that of Europa. Thus, in Fig.~\ref{fig.adm compare} with $\eta_{\rm{melt}}=10^{14}$~Pa~s, $\eta_{\rm{melt}}>\eta_{\rm{Maxwell}}$ for Enceladus while $\eta_{\rm{melt}}<\eta_{\rm{Maxwell}}$ for Europa. A lower temperature of Enceladus leads to a steeper viscosity gradient within Enceladus' shell. This reduces the thickness of the layer where viscosity close to $\eta_{\rm{Maxwell}}$, thus causing an overall reduction of the shell tidal heating. As a result, tidal heating does not affect the shell temperature profile on Enceladus as strongly as it does for Europa.

For the bottom phase boundary, Fig.~\ref{fig.adm compare} shows lower values of admittance compared to the bottom material boundary at low degrees for our Enceladus model. In general, Airy-compensated values of admittance only hold for low spherical harmonic degrees, or long wavelength is much greater than the shell thickness. Thus, we reach the same conclusion as \cite{vcadek2019airy} that the Airy isostasy model can be applied for long wavelengths but may be inaccurate at short wavelengths for the bottom phase boundary.

\subsection{Sensitivity of future gravity and shape data to the icy shell structure}

In this subsection, we characterize what can be achieved with future data at Europa and Enceladus by comparing various shell structure endmembers to expected mission performance. To access the accuracy of admittance recovery two items are needed. First, one needs to estimate the accuracy of gravity coefficients determination, which can be done either with a simplified approach of \cite{bills2019simple} or with a full mission simulation using covariance analysis. Second, a global shape model or, at least, an estimate of the shape power spectrum is required. A global shape model is available for Enceladus \citep{tajeddine2017true} but is not currently available for Europa to our knowledge. Thus, we can only approximately judge about admittance recovery at Europa based on the expected resolution of the gravity field.

\cite{park2011detecting} conducted a covariance analysis study simulating Europa Clipper flyby mission and found that Europa's gravity field may be resolved up to degree~10. A transition between low (Airy compensation-like) admittance values to higher, uncompensated values marks the shift of the dominant topography support mechanism from buoyancy to viscous stresses within the shell. Thus, we can pose a question: under what conditions does this transition occur at $n<10$? We find the transition is captured in admittance spectra at $n<10$ for shell thicknesses over $\approx$24~km for $\eta_{\rm{melt}}=10^{14}$~Pa~s (Fig.~\ref{fig.adm thickness}A). If a lower melt viscosity is assumed ($\eta_{\rm{melt}}=10^{13}$~Pa~s, not shown in Fig.~\ref{fig.adm thickness}A), the tidal heating is stronger within the shell and the admittance transition is captured for shell thicknesses over $\approx$15~km. Thus, we conclude, that admittance is more sensitive to warm and thick shells. Extending the capability of capturing the transition of topography support mechanism to higher degrees will improve the sensitivity of admittance to thinner shells.

\cite{Ermakov2021enceladusgeophys} presented a covariance analysis for Enceladus orbiter missions. \cite{Ermakov2021enceladusgeophys} studied the recovery of Enceladus' gravity field and tides simulating one month of continuous radio-tracking for a single orbiter with radio-tracking to the Earth and a GRAIL-like dual spacecraft with inter-satellite tracking. The ranging accuracy was assumed $10^{-7}$~km~$\rm{s}^{-1}$ for the single spacecraft case and $10^{-9}$~km~$\rm{s}^{-1}$ for the dual spacecraft case. These accuracies are typical for X-band and Ka-band ranging, respectively. We used the covariance analysis by \cite{Ermakov2021enceladusgeophys} and Enceladus shape model by \cite{tajeddine2017true} to estimate the performance of two orbiter mission configurations in recovering admittance. 
The gravity error RMS spectra for the two orbiter configurations are found from the diagonal elements (i.e., variances) of the gravity covariance matrix $\sigma_{\bar{C}_{nm}}^2$ and $\sigma_{\bar{S}_{nm}}^2$ as:

\begin{equation}
    M_n^{gg} = \sqrt{\frac{\sum_{m=0}^n (\sigma_{\bar{C}_{nm}}^2+\sigma_{\bar{S}_{nm}}^2)}{2n+1}}.
    \label{eq: gravityRMS}
\end{equation}

\begin{figure}
    \centering
    \includegraphics[width=\columnwidth]{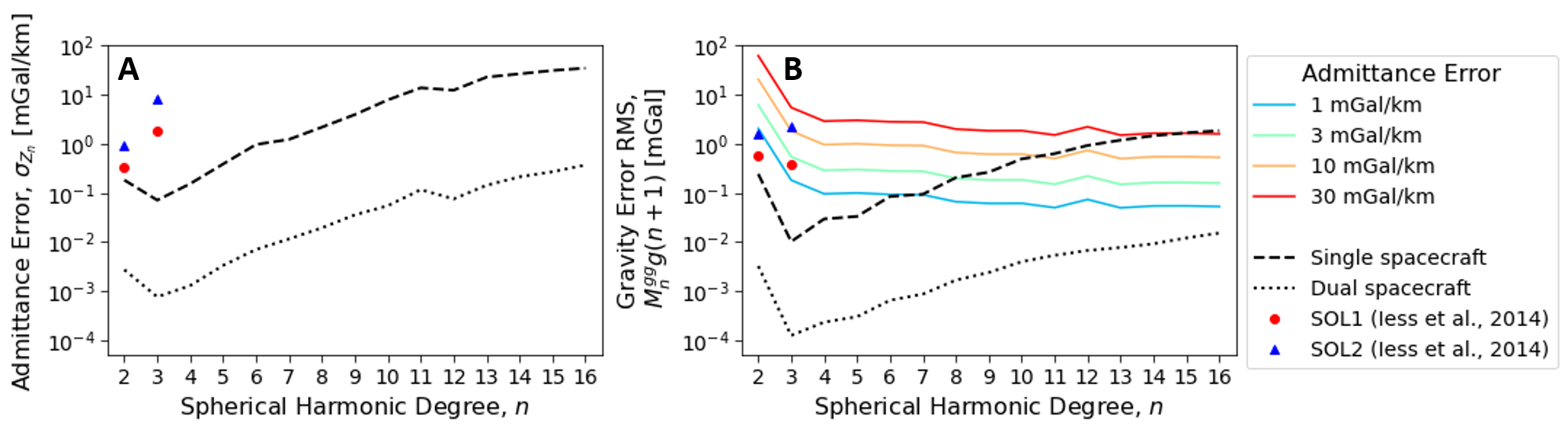}
    \caption{Comparison of different admittance error thresholds to simulated observational error from two Enceladus orbiter missions from \cite{Ermakov2021enceladusgeophys} and current gravity error from \cite{iess2014gravity}. Admittance error ({\bf{A}}) and gravity error RMS spectra ({\bf{B}}) are shown for a single spacecraft with X-band tracking with $10^{-7}$~km~$\rm{s}^{-1}$ accuracy and a GRAIL-like dual spacecraft with inter-satellite Ka-band tracking with $10^{-9}$~km~$\rm{s}^{-1}$ accuracy (corresponding to Fig. 6 in \cite{Ermakov2021enceladusgeophys}). Current observational gravity error and admittance error from two gravity field models (SOL1 and SOL2) are shown for degrees~2 and 3 \citep{iess2014gravity}. A set of gravity performance curves is shown for a given admittance error threshold with factor $g(n+1)$ with gravitational acceleration of Enceladus $g$ for units of mGal ({\bf{B}}).}
    \label{fig.adm error}
\end{figure}

\noindent The gravity error RMS spectra for the two orbiter configurations are shown in Fig.~\ref{fig.adm error}B and correspond to blue and yellow curves in Fig. 6 in \cite{Ermakov2021enceladusgeophys}. In order to match \cite{Ermakov2021enceladusgeophys}, we present the error in radial gravitational acceleration coefficient $M_n^{gg} g(n+1)$, where $g$ is surface gravity. For comparison, we show gravity error RMS from two currently available gravity field models by \cite{iess2014gravity} for degrees~2 and 3. The variance of admittance at degree $n$ is given by:

\begin{equation}
    \sigma_{Z_n}^2 = \mathbf{D}_n \mathbf{C}_n \mathbf{D}_n^T, \\
\label{eq: admittance error}
\end{equation}

\noindent where $\mathbf{D}_n$ is a vector of partial derivatives of degree-$n$ admittance with respect to gravity coefficients:

\begin{equation}
    \mathbf{D}_n = \frac{\bar{\bf{h}}_n}{V_n^{tt}} \frac{G M}{R^3} (n+1).
\end{equation}

\noindent Here, $V_n^{tt}$ is the shape variance spectrum (see Eq. \ref{eq: variance/power spectra} in Appendix \ref{apdx admittance} for definition),  $\bar{\bf{h}}_n = [\bar{A}_{n0},\bar{A}_{n1},\bar{B}_{n1},...,\bar{A}_{nn},\bar{B}_{nn}]$ is a vector of normalized shape coefficients from the shape model of \cite{tajeddine2017true}, and $\mathbf{C}_n$ is the degree $n$ submatrix of the covariance matrix from \cite{Ermakov2021enceladusgeophys}. Note that \cite{tajeddine2017true} provided un-normalized shape coefficients. Shape is assumed to be exact, thus the error comes solely from gravity. The admittance error $\sigma_{Z_n}$ is shown for the single and dual spacecraft configurations in Fig.~\ref{fig.adm error}A. Excluding covariances by looking at the diagonal elements of $\mathbf{C}_n$, we find the correlations between recovered coefficients contribute little to the admittance error for both orbiter configurations. Similarly, admittance error is shown for the gravity error for the two gravity field models from \cite{iess2014gravity}. 

The ability to distinguish between different endmembers of the shell structure using admittance can be used to place a requirement on the gravity error. A gravity error RMS spectum is estimated from an admittance error requirement $\sigma_{Z_n}$ by:

\begin{equation}
    M_n^{gg} = \sigma_{Z_n} \sqrt{V_n^{tt}} \left( \frac{G M}{R^3} (n+1) \right)^{-1},
\end{equation}

\noindent where we assume gravity error uniform across all orders for each degree and there are no correlations between coefficients of different orders. A range of admittance errors from 1 to 30 mGal $\rm{km}^{-1}$ are shown as gravity error RMS spectra in Fig.~\ref{fig.adm error}B. An admittance error requirement is satisfied if the gravity error RMS of a given spacecraft mission configuration is below that for a desired admittance error represented by colored curves Fig.~\ref{fig.adm error}B). For the model of Enceladus shown in Fig.~\ref{fig.adm compare}, the difference in admittance due to the boundary conditions are from $\approx$20 to 30 mGal $\rm{km}^{-1}$ at degrees up to 10. As can be seen in Fig.~\ref{fig.adm error}B, both mission configurations studied in \citep{Ermakov2021enceladusgeophys} would yield admittance accuracy smaller than the expected difference between the two boundary conditions. Thus, such mission configurations would provide sufficient accuracy to distinguish between two types of behaviour of the shell-ocean interface. 

The use of admittance may augment ice-penetrating radar in constraining shell thickness. One of the goals of Europa Clipper's REASON (Radar for Europa Assessment and Sounding: Ocean to Near-surface) instrument is to detect the ice-ocean interface. However, the attenuation of radio waves within ice limits the detection of the ice-ocean interface for the case of a thick shell. \cite{kalousova2017radar} find that direct detection of Europa's ocean with REASON may be possible up to 15~km for a conductive shell or in an area of cold downwelling within a convective shell. In the case that the icy shell of Europa is thick, admittance, which we find is more sensitive to thicker shells, may augment radar in the constraining of shell thickness. 

\subsection{Sensitivity of admittance to tidal dissipation}
\label{sec. sensitivity}

\begin{figure}
    \centering
    \includegraphics[width=\columnwidth]{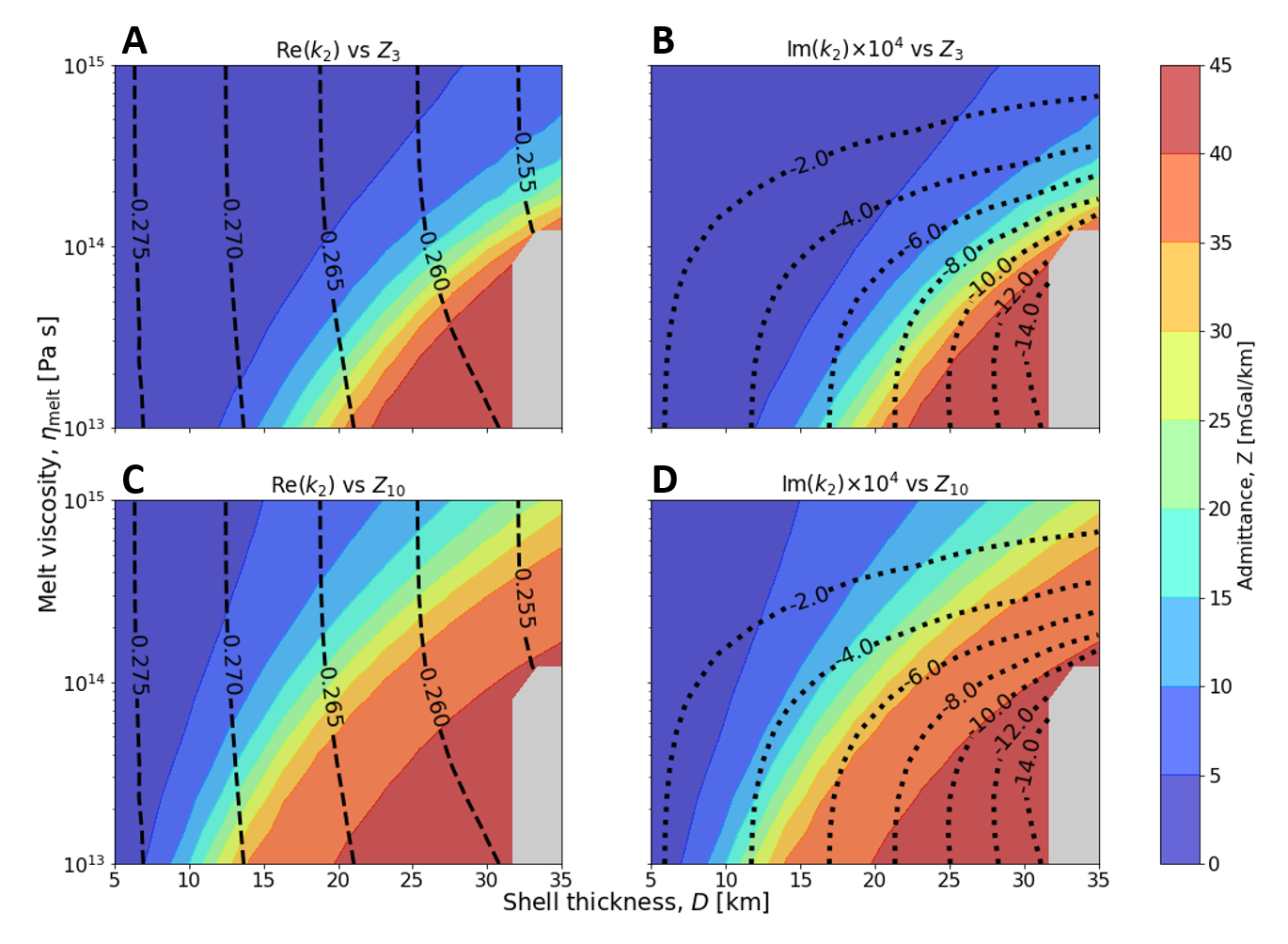}
    \caption{A comparison of real and imaginary parts of the degree 2 tidal Love number, $k_2$ to the values of admittance for Europa computed for a range of shell thickness $D=5-30$ km and melt viscosity $\eta_{\rm{melt}}=10^{13}-10^{15}$ Pa s using the material boundary condition at the base of the shell. Contours of $\mathrm{Re}(k_2)$ (dashed) are plotted on panels ({\bf{A}}) and ({\bf{C}}). Contours of $\mathrm{Im}(k_2)$ (dashed) are plotted on panels ({\bf{B}}) and ({\bf{D}}). Filled contours of admittance at degree 3 are shown on panels ({\bf{A}}) and ({\bf{B}}). Filled contours of admittance at degree 10 are shown on panels ({\bf{C}}) and ({\bf{D}}). A viscosity bound of $10^{24}$~Pa~s was applied. Combinations of thickness and $\eta_{\rm{melt}}$ where a temperature profile did not converge according to our criteria (Section \ref{sec. temperature}) are shown in grey.}
    \label{fig.contour}
\end{figure}

A measurement of the tidal phase lag, or equivalently, the imaginary part of the tidal Love number $k_2$ can provide a constraint on the total dissipation within the satellite \citep{peale1979melting}. However, this measurement is challenging and so far has been achieved only for the Earth \citep{ray2001constraints}, the Moon \citep{williams2014lunar} and Mars \citep{bills2005improved}, which have abundant data sets. Early future data for ocean worlds will likely be more limited. \cite{park2011detecting} estimated that Europa Clipper will enable determination of $k_2$ of Europa with an uncertainty of 0.009. A more recent study by \cite{verma2018expected} is more pessimistic and predicts that Europa's $k_2$ can be determined with an uncertainty of $\approx0.05-0.06$ depending on the Europa Clipper trajectory and the selection of the ground-based radio-tracking assets. We computed the real and imaginary parts of Europa's Love numbers (See Appendix \ref{apdx tidal} for details) and explored their dependence on the shell thickness and viscosity profile. Fig.~\ref{fig.contour} shows how $\mathrm{Re}(k_2)$ and $\mathrm{Im}(k_{2})$ depend on the shell thickness and melt viscosity $\eta_{\rm{melt}}$ for a model of Europa with viscosity bound $\eta_{\rm{bound}} = 10^{24}$~Pa~s. We observe that $\mathrm{Re}(k_2)$ depends primarily on the shell thickness. On the other hand $\mathrm{Im}(k_{2})$, and therefore total dissipation, varies strongly depending on both shell thickness and $\eta_{\rm{melt}}$. In addition, Fig.~\ref{fig.contour} shows gravity-topography admittance at degree 3 and 10 as filled contours. It can be seen that the admittance contours are, in general, not parallel to the contours of $\mathrm{Re}(k_2)$ or $\mathrm{Im}(k_2)$. Thus, measuring admittance would provide non-degenerate information about the dissipation within the icy shell. 

We note that the values of $\mathrm{Im}(k_2)$ are smaller than the estimated uncertainty of $k_2$ recovery \citep{park2011detecting,verma2018expected} assuming $\mathrm{Im}(k_2)$ is measured to the same precision as $\mathrm{Re}(k_2)$. Thus, directly constraining total tidal dissipation within Europa will be challenging from the Europa Clipper data. \cite{park2011detecting} showed that the gravity field of Europa can be estimated to degree 10. More recently, \cite{verma2018expected} showed that only degree 3 and 4 of the gravity field can be estimated. The global shape can be estimated from fitting limb profiles or from building a geodetic network of reference points, which has been done previously for Enceladus using the \textit{Cassini} flyby data \citep{nimmo2011geophysical,tajeddine2017true}. Thus, since measuring the gravity field, the shape and, therefore, gravity-topography admittance, at low degrees is less demanding than measuring $\mathrm{Im}(k_{2})$, admittance measurements would likely allow constraining tidal dissipation within Europa's shell prior to the $\mathrm{Im}(k_{2})$ measurement becoming available. 

Finally, if tidal dissipation is estimated from both $\mathrm{Im}(k_{2})$ and admittance, the potential disagreement between these estimates could indicate either a violation of our modeling assumptions such as: the choice of $\eta_{\rm{bound}}$ and the rheology of ice near the cold surface; the choice of the bottom boundary condition; the assumptions of uniformity of tidal heating distribution and conductive heat transport; or that significant heating occurs in the rocky mantle or the ocean.

\section{Conclusions}

We simulated the viscous relaxation of the icy shells of ocean worlds to test the sensitivity of gravity-topography admittance to the structure of the shell. We find that admittance is sensitive to the topography support mechanism. Our models show that the behavior of the bottom interface---whether it is primarily a material or phase interface---strongly influences the admittance spectrum. The base of the icy shell may behave as a phase boundary at low degrees where tidal heating can drive melting and freezing. At higher degrees, a lack of a heat source would render a material boundary condition more applicable. If the base is treated as a material boundary, low viscosities cause the relaxation of basal topography by lateral flow faster than the relaxation of higher-viscosity surface topography, leading to uncompensated topography. This makes the Airy isostasy model not suitable for interpreting admittance at degrees for which bottom material boundary is applicable except at the longest scales, where topographic loads with wavelengths much larger than shell thickness do not sense the viscosity variation throughout the shell.

A transition between topography support mechanisms may be resolved with future measurements, allowing us to probe the shell structure of ocean worlds. The viscosity structure of the shell is controlled by the rheology of ice. In particular, the poorly constrained grain size of ice influences the tidal heat production and the extent of the low viscosity region at the base of the shell affecting the temperature profile. At low degrees, admittance spectra are more sensitive to temperature and viscosity structures for thick shells with high tidal dissipation. Measurements of higher-degree admittance would allow for studying a wider range of potential shell structures. The higher sensitivity of admittance to the thickness of thicker shells may augment constraining the shell thickness by ice-penetrating radar, which is more sensitive to thinner shells. Thus, a combination of radar and gravity measurements in future ocean world missions may improve the robustness of the measurement strategy. 

We also find that admittance measurement can be used to constrain the tidal dissipation within the icy shell. Such a measurement would be complementary to a demanding measurement of the imaginary part of the tidal Love number, which is proportional to the total tidal dissipation. Measuring admittance in addition to $\mathrm{Im}(k_{2})$ can be used to separate the shell heating from the mantle and ocean heating. Knowledge of heating distribution within ocean worlds would allow for a better understanding of the heat budget and stability of their oceans, which is critical for their long-term habitability.


\acknowledgements
This work sprung form conversations with Roger Fu. The authors thank Ryan Park for providing covariance matrices for gravity measurements from simulations. The authors also thank Peter James, Doug Hemingway, Mikael Beuthe, and Shunichi Kamata for their helpful insights. RA was supported by the National Science Foundation-Department of Energy (DOE) partnership for plasma science and engineering (grant DE-SC0016248). AE received support through a Cassini Data Analysis grant (NNX16AI43G). BM acknowledges support from the U.S. Department of Energy and National Nuclear Security Administration (grant DE-NA0003842).

\section*{Appendix}
\renewcommand{\thesubsection}{\Alph{subsection}}
\renewcommand{\theequation}{\thesubsection.\arabic{equation}}
\numberwithin{equation}{subsection}

\subsection{Tidal heating}
\label{apdx tidal}

We compute the volumetric tidal heat production rate using a spherically symmetric internal structure model and a Maxwell model of viscoelasticity. Our modeling follows the steps of \cite{takeuchi1972seismic}, who define six radial functions $y_i^n(r)$ to describe tidal flow field. These functions are radial multipliers of the corresponding fields expanded in spherical harmonics. Index $n$ refers to a spherical harmonic degree. We restrict our analysis to degree 2 tides and, for now on, we will omit the dependence on $n$. The functions $y_i$ describe radial and tangential displacements ($y_1$ and $y_3$, respectively) and radial and tangential stresses ($y_2$ and $y_4$) expanded in vector spherical harmonics. Gravitational potential ($y_5$) is expanded in scalar spherical harmonics. $y_6$ contains the radial derivative of gravitational potential and is formulated in the following way by \cite{takeuchi1972seismic} to simplify the surface boundary condition:

\begin{equation}
	y_6(r) = \frac{d y_5(r)}{d r} - 4 \pi G \rho y_1(r) + \frac{n+1}{r} y_5(r),
	\label{eq. y_6}
\end{equation}

\noindent where $\rho$ is the density of the layer. Note that definition for $y_6(r)$ is different from another commonly used notation of \cite{alterman1959oscillations}. These six radial functions are found by solving a system of linear differential equations within layers of constant density and viscoelastic parameters:

\begin{equation}
	\frac{d \mathbf{y}(r)}{d r} = \mathbf{A}^{\rm{tidal}} \mathbf{y}(r),
	\label{eq. TS}
\end{equation}

\noindent where $\mathbf{A}^{\rm{tidal}}$ is a $6\times6$ matrix given by Eq. 82 in \cite{takeuchi1972seismic}. The surface boundary condition at $r=R$ for surface radius $R$ are given by vanishing stresses ($y_2(R)=y_4(R)=0$) and $y_6(R)=(2n+1)/R$ describing potential and its derivative continuity. Across the liquid-solid interfaces, which occurs at the ice-ocean and ocean-mantle boundaries for our three-layer model, additional boundary conditions are required. At the center of the planet $r=0$, displacements are zero ($y_1(0) = y_3(0) = 0$) and gravitational potential is also zero ($y_5(0)=0$). The system of differential equations becomes singular at $r=0$. An analytical solution at the surface of a small homogeneous sphere is used as a starting solution of radial functions $y_i(r)$ (see \cite{takeuchi1972seismic} and \cite{martens2016using} for more detail). This starting solution at $r_0 \ll R$, consisting of three linearly independent solutions, is then propagated through successive layers by solving Eq. \ref{eq. TS} with an eighth order Runge-Kutta numerical integrator scheme using variable normalization from \cite{martens2016using}. Finally, the three independent solutions at the surface are linearly weighted to satisfy the surface boundary condition and combined to form the six radial functions $y_i(r)$. Complex-valued tidal Love number $k_2$ is found as $y_5(R)-1$. 

To find the radial distribution of tidal heating, we follow \cite{tobie2005tidal}. Tidal heating is driven by the periodic tidal potential. The degree 2 tidal potential for a tidally-locked satellite on an eccentric orbit, computed to the first order in eccentricity, and evaluated on the surface of the satellite is given by \cite{moore2000tidal}:

\begin{equation}
\Phi = R^2\omega^2 e \left[-\frac{3}{2} P_{2,0} (\cos\theta)\cos \omega t + \frac{1}{4} P_{2,2}(\cos\theta) \left(3\cos\omega t \cos 2\phi + 4\sin\omega t\sin 2\phi \right) \right],
\label{eq. TidalPotential}
\end{equation}

\noindent where $\omega$ is the orbital (and rotational) frequency, $e$ is the eccentricity, $\theta$ is colatitude, $\phi$ is longitude, $t$ is time, $P_{2,0}$ and $P_{2,2}$ are unnormalized associated Legendre functions.

Our internal structure model consists of three main layers. The solid icy shell and mantle are described by the elastic shear modulus $\mu$, Poisson’s ratio $\nu$, and viscosity $\eta$, while the ocean is described by bulk modulus $\kappa$ following the values given in Table \ref{tab. params}. The viscosity profile in the shell is broken up into 80 constant viscosity layers. The viscoelastic Maxwell rheology moduli used in solving for $y_i(r)$ are given by complex-valued Lam\^{e}’s first parameter $\tilde\lambda$ and complex shear modulus $\tilde\mu$ in terms of their real-valued counterparts along with the bulk modulus $\kappa$ and viscosity $\eta$:

\begin{gather}
    \tilde\lambda = \frac{\omega\lambda i + \mu\kappa/\eta}{\omega i + \mu/\eta},\\
    \tilde\mu = \frac{\omega\mu i}{\omega i + \mu/\eta}.
\label{eq. viscoelastic}
\end{gather}

We then use the radial functions $y_i(r)$ to find the sensitivity parameter $H_\mu$ introduced by \cite{tobie2005tidal} to describe the radial sensitivity to the shear modulus (Eq. 33 of \cite{tobie2005tidal}), as it enables computing the radial distribution of the volumetric dissipation rate. A simplified version of $H_\mu$ is given by Eq. 25 in \cite{beuthe2013spatial}). Finally, the radial profile of surface averaged volumetric tidal dissipation rate averaged over moon’s orbit is computed with Eq. 37 of \cite{tobie2005tidal} accounting for a sign correction made by \cite{beuthe2013spatial}:

\begin{equation}
h_{\rm{tide}}(r) = \frac{21}{10} \frac{\omega^5 R^4 e^2}{r^2} H_\mu \mathrm{Im}(\tilde\mu).
\label{eq. htide}
\end{equation}

Our tidal dissipation code was benchmarked by comparing our $y_i(r)$ profiles with values from Fig.~C1 and Fig.~C2 of \cite{kamata2015tidal}. 

\subsection{Surface temperature}
\label{apdx surface temperature}
The surface temperature of our spherically symmetric moon is given by the uniform equilibrium temperature using a fast rotator approximation:

\begin{equation}
    T_{eq} = \left[\frac{F(1-A)}{4 \sigma}\right]^{1/4},
    \label{eq: equlibrium temperature}
\end{equation}

\noindent where $F$ is solar irradiance, $A$ is the Bond albedo, and $\sigma$ is the Stefan-Boltzmann constant. For Europa, we use solar irradiance at Jupiter of 50.26~W~$\mathrm{m}^{-2}$ and Bond albedo 0.68 \citep{grundy2007new} and find surface temperature $T_s\approx 92$~K. For Enceladus, we use solar irradiance at Saturn 14.82~W~$\mathrm{m}^{-2}$ and a Bond albedo of 0.81 \citep{spencer2006cassini} and find surface temperature $T_s\approx 59$~K. 

\subsection{Stokes flow with self-gravitation}
\label{apdx stokes flow}

Viscous relaxation of the icy shell is modeled by solving for the Stokes flow of an incompressible fluid in a spherical shell with self-gravitation. Similarly, to the tidal heating computation, a spherically symmetric icy shell is broken up into layers of constant density $\rho$ and viscosity $\eta$, where the viscosity is set by the temperature profile found in Section \ref{sec. temperature}. The governing equations are:

\begin{equation}
\begin{aligned}
\nabla \cdot v = 0,\\
\nabla^2 V = -4\pi G \rho,  \\
0 = -\nabla p + \eta \nabla^2 v + F,
\label{eq. viscous relaxation}
\end{aligned}
\end{equation}

\noindent where $v$ is the velocity, $V$ is the potential, $p$ is the pressure, $G$ is the universal gravitational constant and $F$ is the external force, equal to gravity in our case. To solve for the Stokes flow, we follow \cite{hager1989constraints}, who define six radial functions $y_i^{n}(r)$ where $n$ is spherical harmonic degree (Eq. 4.11 – 4.18). These functions are not to be confused with the $y$ functions discussed above for the tidal heating problem. Radial and poloidal velocities ($y_1^{n}$ and $y_2^{n}$), radial normal stress ($y_3^{n}$), and poloidal shear stress ($y_4^{n}$) are the coefficients of the corresponding vector fields expanded in vector spherical harmonics. Gravitational potential perturbation ($y_5^{n}$) and its radial derivative ($y_6^{n}$) are coefficients of scalar spherical harmonics. The decoupling of the system of equations for the velocities and stresses from that of the potential is achieved by introducing a new set of variables $u_i^n$ and $v_i^n$. Again, we will drop the dependence on degree $n$ for convenience. The components of the state vectors {\bf{u}} and {\bf{v}} are given in terms of the $y_i$ variables as:

\begin{equation}
\begin{aligned}
u_{1} = y_{1}, \\
u_{2} = y_{2}, \\
u_{3} = r y_{3}/\eta_{0} + \rho r y_{5}/\eta_{0}, \\
u_{4} = r y_{4}/\eta_{0}, \\
v_{1} = \rho_{0} r y_{5}/\eta_{0}, \\
v_{2} = \rho_{0} r^{2} y_{6}/\eta_{0},
\label{eq: u and v}
\end{aligned}
\end{equation}

\noindent where $\rho$ is local density, $\rho_0$ is a reference density, and $\eta_0$ is a reference viscosity chosen close to $\eta_{\rm{bound}}$. The radial derivatives for $u_i$ and $v_i$ can be written in matrix form:

\begin{equation}
\begin{aligned}
\frac{d \mathbf{u}}{d\nu} = \mathbf{A}^{\rm{Stokes}}\mathbf{u}, \\
\frac{d \mathbf{v}}{d\nu} = \mathbf{B}^{\rm{Stokes}}\mathbf{v},
\label{eq: UVRadialDerivatives}
\end{aligned}
\end{equation}

\noindent where $\nu = \log(r/R)$, $R$ is the mean radius of the moon, $\mathbf{A}^{\rm{Stokes}}$ is a $4\times 4$ matrix and $\mathbf{B}^{\rm{Stokes}}$ is a $2\times 2$ matrix given by Eq. 4.33 and 4.34 in \cite{hager1989constraints}. These matrices depend only on viscosity $\eta$ within a layer and spherical harmonic degree $n$. We note that the same system of equations \ref{eq. viscous relaxation} is solved by \cite{hager1979kinematic} as a nearly identical coupled $6\times 6$ system. 

With a choice of boundary conditions at the surface and base of the icy shell, a solution can be propagated from the base to the surface analytically using the propagator matrix method \citep[Eq. 4.43 for the {\bf{u}}-system and Eq. 4.44 for the {\bf{v}}-system in][]{hager1989constraints}.

A free-slip boundary condition is imposed at both boundaries, setting poloidal shear stresses and, correspondingly, $u_{4t}$ and $u_{4b}$ to zero. To account for self-gravity in the boundary conditions for $u_3$, we need to solve the second equation of Eqs. \ref{eq: UVRadialDerivatives} for $v_1$ at the surface and the base of the shell. We solve the {\bf{v}}-system given the shape amplitudes of the top and bottom interface: $h_t$ and $h_b$, respectively. We get the following expressions for $v_{1t}$ and $v_{1b}$:

\begin{equation}
\begin{aligned}
v_{1t} = \frac{4\pi G \rho_0 R}{\eta_0(2n+1)} \left[R h_t \rho_1 + r_b h_b \Delta\rho \left(\frac{r_b}{R}\right)^{l+1} \right], \\
v_{1b} = \frac{4\pi G \rho_0 r_b}{\eta_0(2n+1)} \left[R h_t \rho_1 \left(\frac{R}{r_b}\right)^{-l}+ r_b h_b \Delta\rho \right],
\label{eq: SolvingVSystem}
\end{aligned}
\end{equation}

\noindent where $r_b = R-D$ for shell thickness $D$; $\rho_1$ and $\rho_2$ are the densities of the icy shell and the ocean, respectively, with a density contrast of $\Delta\rho=\rho_2-\rho_1$. Using the solution for $v_{1t}$ and $v_{1b}$, we find the boundary conditions for $u_3$ at the surface ($u_{3t}$) and base ($u_{3b}$) using Eq. 4.53 and 4.54 in \cite{hager1989constraints}. The boundary conditions for $u_{3t}$ and $u_{3b}$ are given as functions of shape amplitudes $h_t$ and $h_b$ as:

\begin{equation}
\begin{aligned}
u_{3t} = u_{3}(R) = \left( -\frac{\rho_1 R g_1}{\eta_0} + \frac{4\pi G\rho_1^2 R^2}{\eta_0+2\eta_0 n} \right) h_t
+ \left(\frac{4 \pi G \rho_1 \Delta\rho r_b^2 (r_b/R)^n}{\eta_0+2\eta_0 n}\right) h_b,\\
u_{3b} = u_{3}(R-D) = \left(-\frac{4\pi G \rho_1 \Delta\rho R r_b (r_b/R)^n}{\eta_0+2\eta_0 n}\right) h_t + 
\left( \frac{\Delta\rho r_b g_2}{\eta_0} - \frac{4\pi G \Delta\rho^2 r_b^2 }{\eta_0 + 2\eta_0 n}\right) h_b,
\label{eq: u_3BC}
\end{aligned}
\end{equation}

\noindent where $g_1$ and $g_2$ are the gravitational accelerations at the top and bottom boundary, respectively. We can write these two equations in a matrix form:

\begin{gather}
\begin{bmatrix}
u_{3t} \\ u_{3b} 
\end{bmatrix}=
\begin{bmatrix}
A_t & B_t \\ A_b & B_b
\end{bmatrix}
\begin{bmatrix}
h_t \\ h_b
\end{bmatrix}.
\label{eq: u3_matrix}
\end{gather}

\noindent The propagator matrix solution for the {\bf{u}}-system in Eqs. \ref{eq: UVRadialDerivatives} is:

\begin{equation}
\mathbf{u}_t = \begin{bmatrix}
u_{1t}\\
u_{2t}\\
u_{3t}\\
u_{4t}
\end{bmatrix} = \mathbf{P} \cdot \mathbf{u}_b = \mathbf{P} \cdot \begin{bmatrix}
u_{1b}\\
u_{2b}\\
u_{3b}\\
u_{4b}
\end{bmatrix},
\label{eq: u solution}
\end{equation}

\noindent where $\mathbf{u}_t$ and $\mathbf{u}_b$ are the state vectors at the surface and the base, respectively. $\mathbf{P}$ is the propagator matrix, which is found by a product of layer-wise propagator matrices for each layer of uniform viscosity between bottom ($k$-th) boundary $r_b = r_k = R-D$, and the surface $r_1 = R$.

\begin{equation}
 \mathbf{P} = 
    \begin{bmatrix}
    p_{1,1} & p_{1,2} & p_{1,3} & p_{1,4} \\
    p_{2,1} & p_{2,2} & p_{2,3} & p_{2,4} \\
    p_{3,1} & p_{3,2} & p_{3,3} & p_{3,4} \\
    p_{4,1} & p_{4,2} & p_{4,3} & p_{4,4} \\
    \end{bmatrix} = \mathbf{P}_{1}(r_1,r_{2}) \cdot \mathbf{P}_{2}(r_{2},r_{3})\cdot ... \cdot \mathbf{P}_{k-1}(r_{k-1},r_k)
\label{eq: PropagatorMatrix}
\end{equation}

\noindent  The propagator matrix for the $i$-th layer is given by the following matrix exponential: 

\begin{equation}
    \mathbf{P}_i(r_{i},r_{i+1}) = \exp\left(\mathbf{A}_i^{\rm{Stokes} }\log\left(\frac{r_{i}}{r_{i+1}}\right)\right)
    \label{eq: propagator matrix}
\end{equation}

For the bottom material boundary, we solve for a time-evolving system illustrated in Fig.~\ref{fig. bc}. Thus, the radial flow velocities $u_{1t}$ and $u_{1b}$ are unknowns. We rearrange Eq. \ref{eq: u solution} and use the boundary condition for $u_3$ from Eq. \ref{eq: u_3BC} to express a system of linear differential equations for the flow velocities $u_1$ and $u_2$ at the top and the base of the shell:

\begin{gather}
\begin{bmatrix}
1&0&-p_{1,1}&-p_{1,2}\\
0&1&-p_{2,1}&-p_{2,2}\\
0&0&-p_{3,1}&-p_{3,2}\\
0&0&-p_{4,1}&-p_{4,2}
\end{bmatrix}
\begin{bmatrix}
u_{1t} = \frac{d h_t}{d t} \\ u_{2t} \\ u_{1b} = \frac{d h_b}{d t} \\ u_{2b}
\end{bmatrix}
=
\begin{bmatrix}
0 &p_{1,3}\\
0 &p_{2,3}\\
-1&p_{3,3}\\
0 &p_{4,3}
\end{bmatrix}
\begin{bmatrix}
A_t & B_t \\ A_b & B_b
\end{bmatrix}
\begin{bmatrix}
h_t \\ h_b
\end{bmatrix},
\label{eq: ODEMaterialBC}
\end{gather}

\noindent where $A_{t}$, $A_{b}$, $B_{t}$, $B_{b}$ are the elements of the $u_3$ matrix from Eq. \ref{eq: u3_matrix}. From Eq. \ref{eq: ODEMaterialBC}, we get the time evolution of {\bf{h}} as a sum of two decaying exponentials, resulting in Eq. \ref{eq. shell evolution}. The degree-dependent shape ratio $w_n=-h_b/h_t$ needed to compute admittance converges to the ratio of the elements of the eigen-vector corresponding to the longer relaxation time.

The bottom phase boundary condition is different in that we solve for a dynamic equilibrium illustrated in Fig.~\ref{fig. bc}. Thus, the topography amplitude vector {\bf{h}} is constant in time. This gives us the condition that radial velocity at the surface ($u_{1t}$) is zero. However, although topography at the base $h_b$ is constant, radial velocity ($u_{1b}$) can be nonzero because the flow across the boundary is assumed to be balanced by phase transitions (melting or freezing). Similar to bottom material boundary, $u_{3t}$ and $u_{3b}$ are taken as functions of $h_t$ and $h_b$ using Eq. \ref{eq: u3_matrix} and a free-slip boundary condition gives $u_{4t} = u_{4b} = 0$. We again rearrange the terms in the propagator matrix solution (Eq. \ref{eq: u solution}), expressing a vector of unknowns in terms of the known quantities:

\begin{gather}
\begin{bmatrix}
p_{1,3} B_b     & p_{1,1} & 0 & p_{1,2} \\
p_{2,3} B_b     & p_{2,1} &-1 & p_{2,2} \\
p_{3,3} B_b-B_t & p_{3,1} & 0 & p_{3,2} \\
p_{4,3} B_b     & p_{4,1} & 0 & p_{4,2} 
\end{bmatrix}
\begin{bmatrix}
h_b\\u_{1b}\\u_{2t}\\u_{2b}
\end{bmatrix}
= 
\begin{bmatrix}
0 & -p_{1,3} \\
0 & -p_{2,3} \\
1 & -p_{3,3} \\
0 & -p_{4,3} 
\end{bmatrix}
\begin{bmatrix}
A_t h_t \\ A_b h_t
\end{bmatrix}
\label{eq: phase boundary solution}.
\end{gather}

\noindent We set $h_t=1$ to get four equations with four unknowns and solve for $h_b$, $u_{1b}$, $u_{2t}$, and $u_{2b}$. Finally, we find the shape ratio $w_n = -h_b/h_t = -h_b$, which we use in computing the gravity-topography admittance. 

\subsection{Gravity-topography admittance}
\label{apdx admittance}

In this section, we present how gravity-topography admittance is~1)~computed from the shape and gravity data, and~2)~modeled using the degree-dependent shape ratio $w_n$ found in the previous subsection. Our approach is similar to  \cite{ermakov2017constraints} in their study of Ceres' admittance. The measured shape of the icy moon can be expressed as a spherical harmonic expansion \citep{wieczorek2007gravity}:

\begin{equation}
r(\theta,\phi) = R \left[ 1 + \sum_{n=1}^{\infty} \sum_{m=0}^{n} (\bar A_{nm} \cos(m\phi) + \bar B_{nm} \sin(m\phi))\bar P_{nm}(\cos(\theta)) \right] \label{eq: shape},
\end{equation}

\noindent where $\theta$ is colatitude, $\phi$ is longitude, $R$ is the mean radius, $\bar A_{nm}$ and $\bar B_{nm}$ are normalized coefficients of the spherical harmonic expansion, and $\bar P_{nm}$ are normalized associated Legendre functions for spherical harmonic degree $n$ and order $m$. Similarly, the gravitational potential is also expanded in spherical harmonics:

\begin{equation}
U(r,\theta,\phi) = \frac{G M}{r} \left[ 1 + \sum_{n=2}^{\infty} \sum_{m=0}^{n} \left(\frac{R_0}{r}\right)^n (\bar C_{nm} \cos(m\phi) + \bar S_{nm} \sin(m\phi)) \bar P_{nm}(\cos(\theta)) \right] \label{eq: gravity},
\end{equation}

\noindent where $R_0$ is the reference radius, $r$ is the observation radius, and $\bar C_{nm}$ and $\bar S_{nm}$ are normalized spherical harmonic coefficients of the gravitational potential. Note that the reference radius is not necessarily equal to the mean radius $R$. The variance spectrum of the shape $V_n^{tt}$, gravity $V_n^{gg}$ and the cross-variance spectrum of gravity and shape $V_n^{gt}$ are found from the spherical harmonic coefficients in Eq. \ref{eq: shape} and \ref{eq: gravity} as:

\begin{equation}
\begin{aligned}
V_n^{tt}  =  \sum_{m=0}^{n} (\bar A_{nm}^2+ \bar B_{nm}^2), \\
V_n^{gg}  =  \sum_{m=0}^{n} (\bar C_{nm}^2+ \bar S_{nm}^2), \\
V_n^{gt}  = \sum_{m=0}^{n} (\bar A_{nm} \bar C_{nm} + \bar B_{nm}\bar S_{nm}).
\label{eq: variance/power spectra}
\end{aligned}
\end{equation}

\noindent We are interested in finding the gravity-topography admittance between radial gravitational acceleration and the shape. Thus, we need to differentiate Eq. \ref{eq: gravity} with respect to $r$, which results in simply multiplying each term in Eq. \ref{eq: gravity} by $(n+1)/r$. Gravity-topography admittance $Z_n$ in terms of gravity-shape cross variance and shape variance is given as:

\begin{equation}
Z_n = \frac{V_n^{gt}}{V_n^{tt}}\cdot\frac{G M}{R^3} (n+1).
\label{eq: gravity shape admittance}
\end{equation}

\noindent This expression is known to produce a more accurate admittance estimate in the case when gravity data is more noisy than the shape data \citep{mckenzie1994relationship}, which is typically the case for ocean worlds. It is fully equivalent to:

\begin{equation}
Z_n = \sqrt{\frac{V_n^{gg}}{V_n^{tt}}} R_n^{gt}\cdot\frac{G M}{R^3} (n+1),
\label{eq: gravity_shape_admittance_with_correlation}
\end{equation}

\noindent where $R_n^{gt} = V_n^{gt}/\sqrt{V_n^{gg}V_n^{tt}}$ is the gravity-shape correlation coefficient. We note that in the viscous relaxation models presented in this paper, the gravity and shape are always ``in phase''. Thus, the correlation is either 1 or -1. Since our model concerns small-amplitude topography, it is spectrally pure. That is the topography harmonic of degree $n$ and order $m$ produces the gravity perturbation only of the same degree and order. Thus, Eq. \ref{eq: gravity_shape_admittance_with_correlation} can be simply written in terms of shape and gravity coefficients:

\begin{equation}
Z_n = \frac{\sigma}{h}\cdot\frac{G M}{R^3} (n+1),
\label{eq: GravityTopographyAdmittnaceSimple}
\end{equation}

\noindent where $\sigma=\left[\bar C_{nm}, \bar S_{nm}\right]$ and $h=\left[\bar A_{nm}, \bar B_{nm}\right]$ are degree-$n$ coefficients of gravity and shape, respectively. 

In order to compute the model admittance, we need to find the relationship between the gravity and shape spherical harmonic coefficients. Assuming a homogeneous spherical body with topography small in amplitude relative to its wavelength (a ``mass-sheet'' approximation), we can use the first-order term of Eq. 10 in \cite{wieczorek1998potential} to get the linear relationship between gravity and shape coefficients:

\begin{equation}
\sigma=\frac{3}{2n+1} \cdot \left(\frac{R}{R_{vol}}\right)^3 \cdot  \left(\frac{R}{R_{0}}\right)^n \cdot h,
\label{eq: GravTopoLinearRelationship}
\end{equation}

\noindent where $R_{vol}$ is the radius of the volume-equivalent sphere.

For a body consisting of multiple layers, topography at the interfaces associated with density changes generates perturbation in gravity. We only consider topography at the surface and the base of the icy shell assuming the ocean floor has no topography. Thus, the total gravity includes contributions only from the upper two interfaces. The gravity contribution of each layer needs to be weighted by fractional mass of that layer $(\rho_i/\bar\rho) (r_i/R)^3$, where $\bar\rho$ is the mean density of the body. Since we seek the gravity coefficients referenced to the mean radius of the body ($R_0 = R$ in Eq. \ref{eq: gravity}), the contributions to gravity from $i$-th layer must be upward propagated by $(r_i/R)^n$ to find the contribution to gravity at the surface. Finally, we reference the topography coefficients to the mean radius $R$ of the body in Eq. \ref{eq: shape}. Thus, the topography at the base of the shell $h_b$ needs to be upscaled by $R/(R-D)$. In summary, we find the total gravity coefficient at degree $n$ as:

\begin{equation}
\sigma = \frac{3}{2n+1}\cdot h_t \cdot  \underbrace{ \left( \frac{\rho_1}{\bar\rho} \right) }_{\text{\makebox[0pt]{fractional mass}}}  + \frac{3}{2n+1} \cdot h_b \cdot  \underbrace{\left(\frac{R}{R-D} \right)}_\text{rescaling} \cdot \underbrace{\left[\frac{\rho_2-\rho_1}{\bar\rho} \cdot \left(\frac{R-D}{R}\right)^3\right]}_\text{fractional mass}  \cdot  \underbrace{ \left(\frac{R-D}{R}\right)^n}_{\text{\makebox[0pt]{upward propagation}}}.
\label{eq: unsimplified admittance}
\end{equation}

\noindent In this derived expression, we ignored the factor $(R/R_{vol})^3$, since the difference between $R_{vol}$ and $R$ is small for a small-amplitude topography, which is one of our assumptions anyway. Plugging in $h_b = - w_n h_t$, dividing by the surface shape coefficients $h_t$ and multiplying by $G M R^{-3} (n+1)$ to find the ratio of radial gravitation acceleration to shape coefficients (in mGal $\mathrm{km}^{-1}$), the gravity-topography admittance simplifies to:

\begin{equation}
    Z_n=\frac{GM}{R^3}\cdot\frac{3(n+1)}{(2n+1)}\cdot\left[\frac{\rho_1}{\bar{\rho}}-\frac{\Delta\rho}{\bar{\rho}}\left(\frac{R-D}{R}\right)^{n+2}w_n\right].
    \label{eq: Admittance}
\end{equation}

\noindent The shape ratio $w_n$ is found given a viscosity profile and boundary conditions described in Appendix \ref{apdx stokes flow}. Plugging in $w_n = \rho_1/\Delta \rho$ into Eq. \ref{eq: Admittance}, we recover admittance for the standard Cartesian Airy compensation model:

\begin{equation}
    Z_n=\frac{GM}{R^3} \cdot \frac{3(n+1)}{(2n+1)} \cdot \frac{\rho_1}{\bar{\rho}} \cdot \left[1-\left(\frac{R-D}{R}\right)^{n+2}\right].
    \label{eq: AdmittanceAiry}
\end{equation}

\noindent If $w_n = 0$, we recover admittance for an uncompensated topography:

\begin{equation}
    Z_n=\frac{GM}{R^3} \cdot \frac{3(n+1)}{(2n+1)} \cdot \frac{\rho_1}{\bar{\rho}}.
    \label{eq: AdmittanceUncompansated}
\end{equation}




\bibliography{bibliography}{}

\end{document}